\documentclass{PoS}

\usepackage{booktabs}
\usepackage{graphicx}
\usepackage{subfig}

\title{The Yang-Mills gradient flow and renormalization}

\ShortTitle{The Yang-Mills gradient flow and renormalization}

\author{\speaker{Alberto Ramos}\\
        PH-TH, CERN, CH-1211 Geneva 23, Switzerland\\
        E-mail: \email{alberto.ramos@cern.ch}}
\abstract{In this proceedings contribution we will review the main
  ideas behind the many recent works that apply the gradient flow to
  the determination of the renormalized coupling and the
  renormalization of composite operators. We will pay special
  attention to the continuum extrapolation of flow quantities. 
\vspace{2cm}
\begin{flushright}
CERN-PH-TH-2015-128
\end{flushright}
}

\FullConference{The 32nd International Symposium on Lattice Field Theory\\
                 23-28 June, 2014\\
                 Columbia University New York, NY}

\begin{document}

\section{The Yang-Mills gradient flow}

The basic idea of the Yang-Mills flow consists in introducing an extra
dimension to our gauge fields, that we will call flow time and denote
by $t$ (with dimensions [length$^2$]), and study how fields evolve
according to the flow equation 
\begin{equation}
  \label{eq:flow}
  \frac{d B_\mu(x,t)}{dt} = D_\nu G_{\nu\mu}(x,t)\,,
\end{equation}
with initial condition $B_\mu(x,t=0) = A_\mu(x)$, and $G_{\mu\nu}$
being the field strength 
\begin{equation}
  G_{\nu\mu}(x,t) = \partial_\nu B_\mu(x,t) - 
  \partial_\nu B_\mu(x,t) + [B_\nu(x,t),B_\mu(x,t)] \,.
\end{equation}

Although it has been used in different contexts in our
field~\cite{Narayanan:2006rf,Luscher:2009eq,Lohmayer:2011si}, the 
recent growth of works and interest in the gradient flow have to do 
mainly with two facts: first the well understood properties of
composite operators defined at positive flow time ($t>0$) under
renormalization and its continuum
limit~\cite{Luscher:2010iy,Luscher:2011bx,Luscher:2013cpa}, 
and second with the concrete proposals of using the gradient flow to
give a precise definition of the renormalized
coupling~\cite{Luscher:2010iy}, and as a general tool to renormalize
composite operators~\cite{Luscher:2011bx,Luscher:2013cpa}. In this
contribution I will try to summarize the recent works of the community
in these directions. 
 
The key idea behind most of these applications is that 
the r.h.s. of the flow equation~(\ref{eq:flow}) is nothing but the
gradient of the Yang-Mills action, and therefore the  gauge field becomes
smoother along the flow: the flow smears the gauge field over a
region of radius $\sqrt{8t}$, killing the UV divergences and making
composite gauge invariant observables automatically renormalized.  

For example the energy density, an observable that we will see
throughout all this proceedings contribution, defined as
\begin{equation}
  \langle E(x,t)\rangle = -\frac{1}{2} \langle{\rm
    Tr} G_{\mu\nu}(x,t)G_{\mu\nu}(x,t) \rangle\,,
\end{equation}
is finite for
$t>0$. This is not an accident of this particular observable, or of
leading order perturbation theory. The existence of a continuum limit
(to all orders in perturbation theory) for all gauge invariant
composite operators at positive flow time was  proved
in~\cite{Luscher:2011bx}.  

Since $t^2\langle E(t)\rangle$ is a 
dimension-less renormalized quantity that depends on a scale (given by
$\sqrt{8t}$), it is a natural candidate for scale setting and
to define the renormalized coupling, as we will see in the following
pages. 

The second application that we will cover is related with the behavior
of an operator like $\langle E(t)\rangle$ as $t\rightarrow
0$. The correlation function will become singular in a way that can be
described by an expansion in renormalized operators with singular
coefficients~\cite{Luscher:2011bx} 
\begin{equation}
  E(x,t) = \sum_\alpha c_\alpha(t) \{O^\alpha\}_R(x) + \mathcal O(t)\,,
\end{equation}
where $\{O^\alpha\}_R(x)$ are the set of renormalized operators that
``mix'' with 
$E(t)$, and $ c_\alpha(t)$ are the singular coefficients. This so called
\emph{small flow time expansion}, has recently been used in the context of
thermodynamics, as has been reviewed in this
conference~\cite{Kitazawa:2014uxa}. But the idea of 
the small flow time expansion has a more general scope, and we will
dedicate some time to this matter.

For any of these applications taking the continuum limit of flow
quantities is a fundamental step. The community has recently dedicated
some effort in understanding and controlling the cutoff effects of
flow quantities, and these will also be reviewed here.

\section{The gradient flow coupling}

For the moment, most of the applications of the gradient flow to the
definition of a renormalized coupling make use of the energy (or
action) density at positive flow time
\begin{equation}
  \langle E(x,t)\rangle = -\frac{1}{2} \langle{\rm
    Tr} G_{\mu\nu}(x,t)G_{\mu\nu}(x,t) \rangle\,.
\end{equation}
As we have said in the previous section, $\langle E(x,t)\rangle$ is a
renormalized quantity, and in fact it has been shown
in~\cite{Luscher:2009eq} that $E(x,t)$ has a perturbative expansion
given by 
\begin{equation}
  \label{eq:gsqpt}
  \langle E(t)\rangle = 
  \frac{3}{16t^2 \pi^2} g^2_{\overline{MS}}(\mu)
  \left[1+c_1g^2_{\overline{MS}}(\mu)+   
    \mathcal O(g^4_{\overline{MS}})\right] \,,
\end{equation}
where $c_1$ is a finite coefficient. 

This immediately suggests two
applications~\cite{Luscher:2009eq}. First one can define a 
scale $t_0$ fixed by the condition $t^2_0\langle E(x,t_0)\rangle =
0.3$. This subject was reviewed extensively in the last lattice
conference~\cite{Sommer:2014mea}, and we will not talk more about
scale setting. Second, one can use $t^2\langle E(x,t)\rangle$ to
define a renormalized coupling 
\begin{equation}
  \label{eq:ginfv}
  g_{\rm GF}^2(\mu) = \frac{16\pi^2}{3}t^2\langle
  E(x,t)\rangle\Big|_{t = 1/8\mu^2}\,.
\end{equation}

Although Eq.~(\ref{eq:ginfv}) can be used as it is, it requires the
existence of a window $1/L\ll\mu\ll 1/a$ so that the determination of
$g^2_{\rm GF}(\mu)$ is free of both lattice artifacts and finite
volume effects. This Windowing problem can be solved in a very elegant
way by relating the renormalization scale with the finite size of the
system (finite-size scaling)~\cite{Luscher:1991wu}
\begin{equation}
  \mu = \frac{1}{cL}\,,
\end{equation}
where $c$ is a constant that relates both scales. Usually, in an abuse
of notation, one writes $g^2_{\rm GF}(L)$ for this running
coupling. Since this coupling naturally feels the finite
volume in which it is defined, the coefficients in the
perturbative expansion Eq.~(\ref{eq:gsqpt}) will 
depend on the choice of boundary conditions, as will the
concrete coupling definition. All these coupling
definitions can be written as
\begin{equation}
  \label{eq:g}
  g_{\rm GF}^2(L) = \mathcal N^{-1}t^2\langle
  E(x,t)\rangle\Big|_{\sqrt{8t} = cL}\,,
\end{equation}
where $\mathcal N$ will depend on the boundary conditions and choice
of $c$.

Running coupling schemes have been proposed with
periodic~\cite{Fodor:2012td}, Schr\"odinger
functional (SF)~\cite{Fritzsch:2013je}, twisted (a la
t'Hooft)~\cite{Ramos:2014kla}, and mixed
open-SF~\cite{Luscher:2014kea} boundary conditions. The advantages of
each particular choice for a concrete project have to be evaluated
taking into account the following points
\begin{description}
\item[Cutoff effects] Schemes that break the invariance under
  translations, like the SF or open-SF schemes require further work to
  ensure an $a^2$ scaling towards the continuum. Quantum field
  theories defined in manifolds with boundaries require additional
  boundary counterterms for
  renormalization~\cite{Symanzik:1981wd}. For the concrete case of
  Yang-Mills theories on the lattice this implies that additional
  boundary counterterms have to be introduced to avoid $\mathcal O(a)$
  cutoff effects~\cite{Luscher:1992an}. Moreover, Wilson fermions
  require additional boundary and bulk counterterms to ensure
  $\mathcal O(a^2)$ scaling~\cite{Sint:1993un,Sint:2010eh}.

  On the other hand schemes that preserve the invariance under
  translations, like the ones that use periodic or twisted boundary
  conditions, do not have these extra difficulties.  Automatic $\mathcal
  O(a)$ improvement is guaranteed when working with massless
  Wilson quarks~\cite{Frezzotti:2003ni,Sint:2010eh} without having to
  introduce any counterterms.

\item[Zero modes in perturbation theory] One important characteristic
  of the 
  schemes with either twisted, SF or open-SF boundary conditions is
  that the field configuration that minimizes the action is unique up
  to gauge transformations. On the other hand this is not the case for
  the periodic 
  case. The presence of zero-modes that are not gauge degrees of
  freedom complicates substantially any perturbative computation that
  one wants to address~\cite{GonzalezArroyo:1981vw}. Moreover the
  observable $t^2\langle E(x,t)\rangle$ has a non-analytic expansion
  in $\alpha_{\overline{\rm MS}}$, leading to a non-universal
  $\beta-$function. 

  At this point is important to make clear that this is a problem of
  perturbation theory, and any non-perturbative computation is
  insensitive to this issue. But one should not underestimate the
  usefulness of perturbative computations in order to get insight into
  the full non-perturbative results. Also, we recall that many
  applications of these running coupling schemes need at some point
  to match with pertubation theory. This matching is not only
  more difficult because of the additional complications of the
  perturbative computations, but one might have to 
  do it at higher energy scales than in the case of schemes with a
  universal $\beta-$function.

\item[Critical slowing down] Recently the well known problem of
  topology freezing~\cite{DelDebbio:2004xh,Schaefer:2010hu} that
  affects large volume simulations has been
  found to affect also studies of the running
  coupling~\cite{Fritzsch:2013yxa}. Although
  in~\cite{Fritzsch:2013yxa} a way to overcome this problem is
  proposed, essentially by restricting the measurements to a sector
  of fixed topological charge, the open-SF scheme
  is the only scheme that address this issue in a more simple and
  elegant way by choosing some boundary conditions that allow the
  topological charge to flow in and out of the
  lattice~\cite{Luscher:2014kea}. 
\end{description}

In all these studies a central role is played by the step-scaling
function
\begin{equation}
  \sigma_s(u) = g^2_{\rm GF}(sL)\Big|_{g^2_{\rm GF}(L) = u}\,,
\end{equation}
that measures the change of the coupling when the renormalization
scale is changed by a factor $1/s$. One should note that this
step-scaling function can be computed very easily on the lattice by
changing the lattice size $L/a$ while keeping the bare
parameters fixed.

\subsection{Comparison of the gradient flow coupling and the SF
  coupling} 

Studies of the running coupling using the idea of finite-size scaling
have been done until recently using the so called SF
coupling~\cite{Luscher:1992an}. In this case one imposes SF boundary
conditions on the fields and  measures how the action depends on the
induced background field. Here we will review the differences between
the SF and the GF couplings. 

In
Tab.~\ref{tab:comp} we can see some data~\cite{Fritzsch:2013je} of both
the SF and the GF couplings. These data correspond to $N_f=2$
simulations at constant volume $L_1\sim 0.4{\rm fm}$, and therefore at
constant renormalization scale $\mu$. This is why the values of
$g_{SF}^2(L_1)$ are constant within errors. On the other hand the values
of the GF coupling are not constant within errors. This is a
manifestation of the cutoff effects in the GF coupling, since the
values of $g^2_{\rm GF}(L)$ of some particular ensemble depend not
only on $L$, but also on the lattice spacing $a$. We will talk about
cutoff effects in more detail later, but for the moment it is enough
to say that the relative size of cutoff effects decreases when $c$
increases~\cite{Fritzsch:2013je}. 

\begin{table}
  \centering
          \begin{tabular}{clllllllllll}\toprule
                         $L/a$  &  $6$              &  $8$               &  $10$             &  $12$             &  $16$                 \\
                       $\beta$  &  $5.2638$         &  $5.4689$          &  $5.6190$         &  $5.7580$         &  $5.9631$             \\
            $\kappa_{\rm sea}$  &  $0.135985$       &  $0.136700$        &  $0.136785$       &  $0.136623$       &  $0.136422$           \\
                $N_{\rm meas}$  &  $12160$          &  $8320$            &  $8192$           &  $8280$           &  $8460$         
\\\cmidrule(lr){1-6}
 $\overline{g}^2_{\rm SF}(L_1)$ &  $4.423(75)$      &  $4.473(83)$       &  $4.49(10)$       &  $4.501(91)$      &  $4.40(10)$     \\
\cmidrule(lr){1-6}
      $\overline{g}^2_{\rm GF}(\mu)\, ({ c}=0.3)$ &  $4.8178(46)$ &  $4.7278(46)$  &  $4.6269(47)$ &  $4.5176(47)$ &  $4.4410(53)$     \\
      $\overline{g}^2_{\rm GF}(\mu)\, ({ c}=0.4)$ &  $6.0090(86)$ &  $5.6985(86)$  &  $5.5976(97)$ &  $5.4837(97)$ &  $5.410(12)$      \\
      $\overline{g}^2_{\rm GF}(\mu)\, ({ c}=0.5)$ &  $7.106(14)$  &  $6.817(15)$   &  $6.761(19)$  &  $6.658(19)$  &  $6.602(24)$      \\
           \bottomrule
        \end{tabular}


  \caption{Comparison between the gradient flow coupling $g_{\rm
      GF}^2$ and the Schr\"odinger functional coupling $g^2_{\rm SF}$
    in a set of $N_f=2$ ensembles at constant volume given by
    $L_1=0.4{\rm fm}$~\cite{Fritzsch:2013je}.}
  \label{tab:comp}
\end{table}

Tab.~\ref{tab:comp} also shows that the GF coupling is much more
precise than the SF coupling. We have to take into account that the
numbers 
of the GF coupling were produced with one order of magnitude less
statistics than the ones with the SF coupling. In this sense the
observable used to define the GF coupling has much smaller variance,
and therefore leads to more precise determinations. The data in
Tab.~\ref{tab:comp} also show that this variance increases with the
value of $c$. A more quantitative comparison of the variance of both
coupling definitions can be found in~\cite{Fritzsch:2013hda} were it
is argued that the variance of the GF coupling is around 50-100 times
smaller at these volumes.

An important point is how this variance scales both with the lattice
spacing $a$ and the physical volume $L$. Summarizing the studies made
in~\cite{deDivitiis:1994yz,Fritzsch:2013hda}, we can say that the
variance of the SF coupling diverges as one approaches the continuum
like $1/a$, while the variance of the GF coupling is roughly
independent of $a$ in the scaling region. On the other hand, and as
R. Sommer realized very soon, the variance of the SF coupling behaves
in small volumes (small values of the coupling) as $g_{\rm SF}^8$,
while the GF coupling behaves as $g^4_{\rm GF}$. These different
behaviors are used by the ALPHA collaboration to devise an optimal
strategy to compute the running coupling in
QCD~\cite{Brida:2014joa}. In this approach the SF coupling is used in
the small volumes to match with perturbation theory, and the GF
coupling is used in the larger volumes to 
match with a hadronic scheme.

\section{Cutoff effects and the gradient flow}

On the lattice the gradient flow is defined through the equation
\begin{equation}
  \label{eq:flowlat}
  a^2\partial_t V_\mu(x,t) = -g_0^2 \{T^a\partial_{x,\mu}^a S_{\rm fl}(V)\}
  V_\mu(x,t) \,,  \qquad V_\mu(x,0) = U_\mu(x)  \,,
\end{equation}
where $U_\mu(x)$ are the links associated with the fundamental gauge
field, $V_\mu(x,t)$ are those associated with the flow field, and $S_{\rm
  fl}(V)$ is any lattice action (not necessarily the same used in the
simulation). The 
derivative $\partial_{x,\mu}^a$ has to be understood as
Lie-algebra valued (for more details see for
example~\cite{Luscher:2010iy}). Eq.~(\ref{eq:flowlat}) is simply a
lattice version of the flow equation Eq.~(\ref{eq:flow}), in which the
links are evolved according to the gradient of a lattice action. It
has become clear recently~\cite{Fodor:2014cpa,Ramos:2014kka}, that the
choice of the action that defines the flow, together with the choice 
of action for the simulation of the fundamental gauge field and the
choice of discretization for the observable are the ingredients that
determine the cutoff effects on any flow quantity. 

Step-scaling studies and computations of the running coupling have
only one systematic effect that needs to be taken under control:
taking the continuum limit. In the studies with the SF coupling
it is not strange to use small lattices, including those with
$L/a=4,6$. On the other hand schemes that use the gradient flow
coupling typically use 
larger lattices. This difference does not mean that gradient flow
observables have large cutoff effects \emph{per se}, since the
natural scale that controls the size of cutoff effects for flow
quantities is not the whole lattice ($a/L$), but the region that is
smeared by the flow ($a/\sqrt{8t}=a/cL$). This suggests that larger values
of $c$ might lead to smaller cutoff effects, and this is in fact the
case~\cite{Fritzsch:2013je}. But this comes at a price: first, as we
have seen, couplings 
defined at larger values of $c$ have larger errors. In addition,
in schemes where the lattice has boundaries in the time direction (SF
or open-SF), the boundary effects become more noticeable at larger
$c$~\cite{Luscher:2014kea}\footnote{Note that this can be ameliorated
  by simply increasing the ratio $T/L$.}.  

\begin{figure}
  \centering
  \subfloat[][]{
    \includegraphics[width=0.45\textwidth]{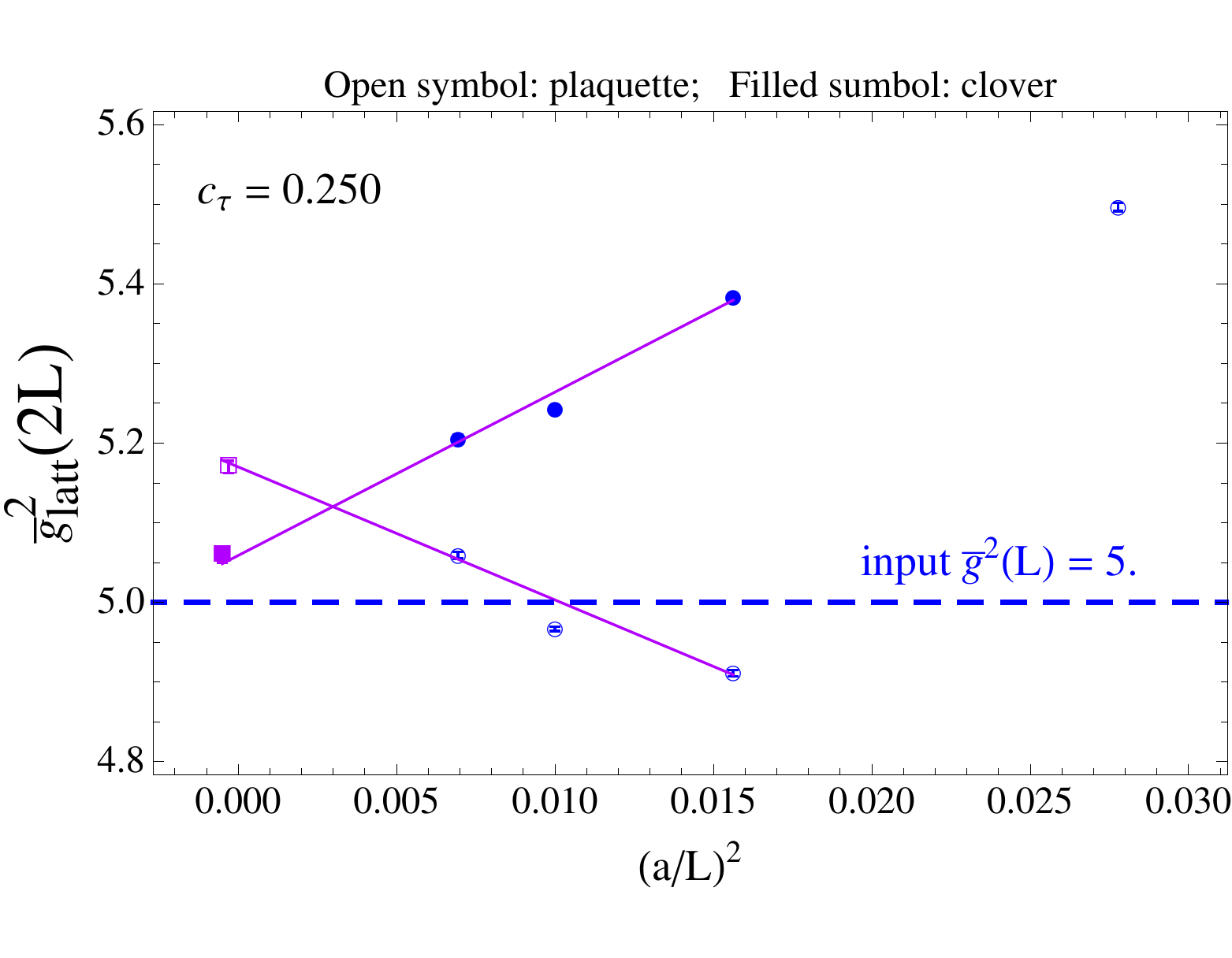}
  }
  \subfloat[][]{
    \includegraphics[width=0.45\textwidth]{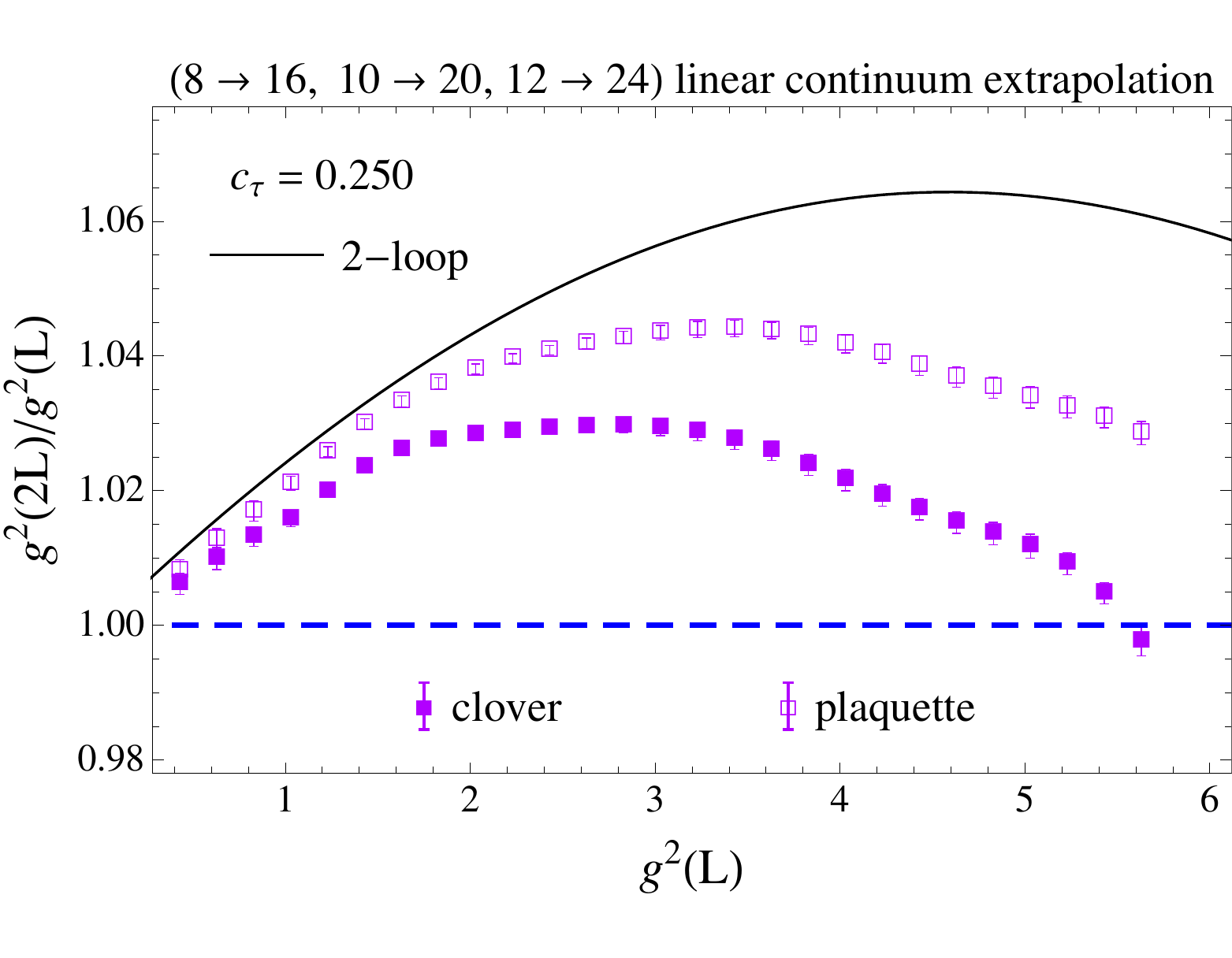}
  } \\
  \subfloat[][]{
    \includegraphics[width=0.45\textwidth]{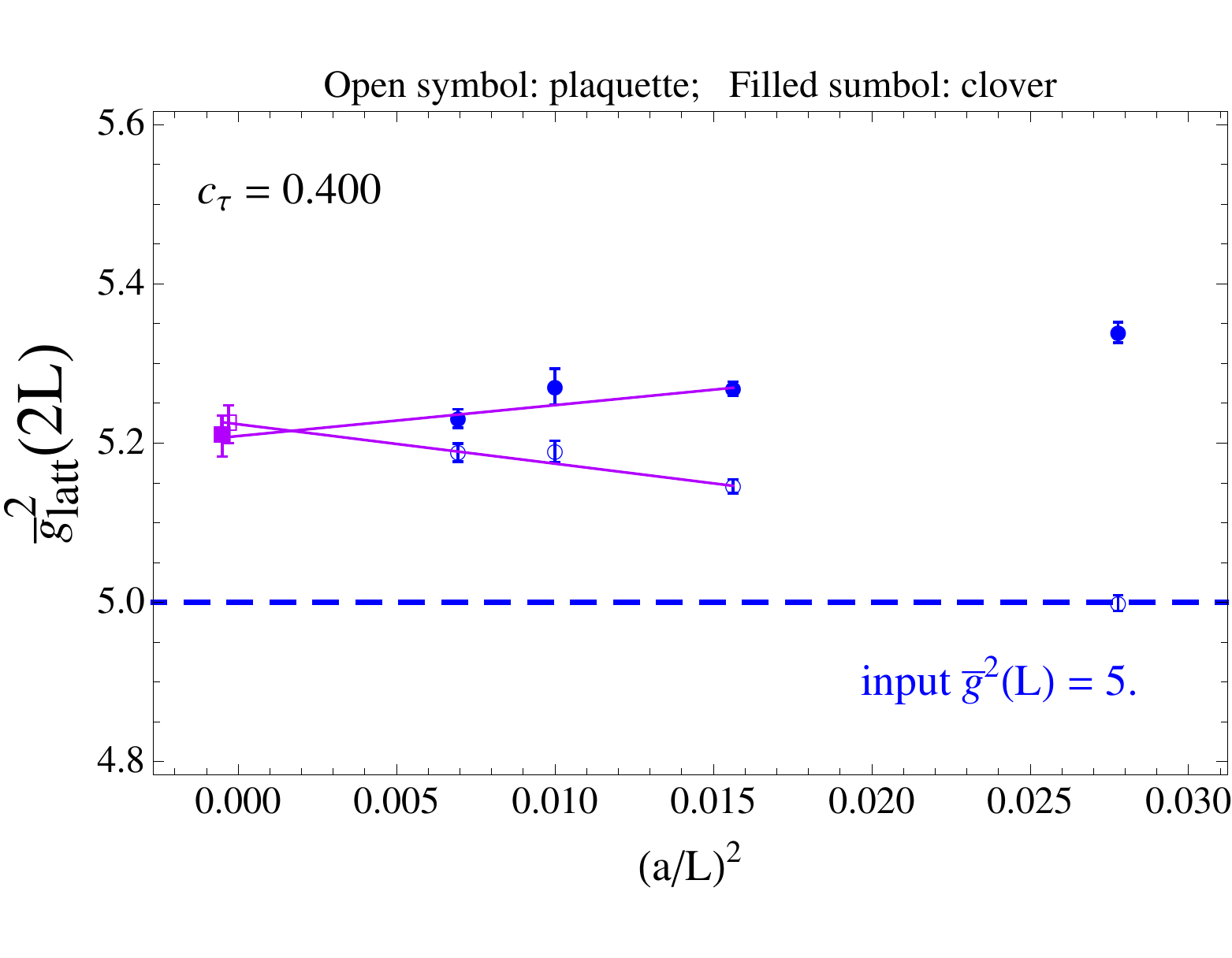}
  }
  \subfloat[][]{
    \includegraphics[width=0.45\textwidth]{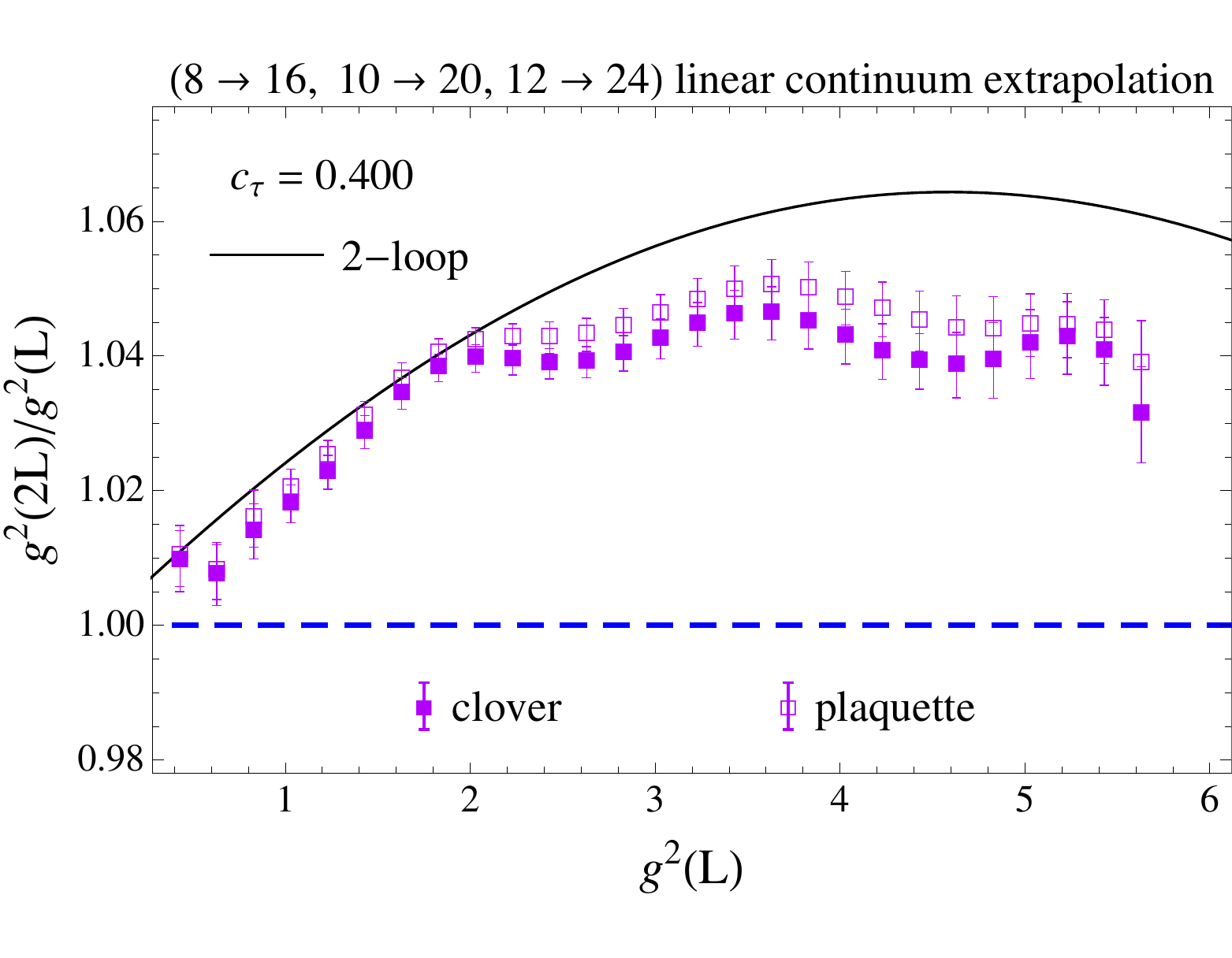}
  } 
  \caption{Continuum extrapolation of the step scaling function
    $\sigma_2(u)$ and running coupling using different discretizations
    for the observable 
    $E(x,t)$. Top row: $c=0.25$, Bottom row $c=0.4$. Left Column:
    Continuum extrapolation for $\sigma_2(5)$. Right column:
    $\sigma_2(u)/u$ in the continuum as a function of
    $u$. See the text for more details. (Source~\cite{Lin:2014fxa}).} 
  \label{fig:cont1}
\end{figure}

In order to illustrate the dangers of the continuum extrapolation
in step scaling studies, I will use some data from these proceedings
contribution~\cite{Lin:2014fxa} (see Fig.~\ref{fig:cont1}). As we can
see, when two different discretizations for the observable $E(x,t)$
are used, the continuum extrapolations of the step scaling function do
not agree for small values of the parameter $c$ (Fig.~\ref{fig:cont1}
(a)).  Note that in this particular case,
if one only looks at one observable and only at one value of $c$, it
would be difficult to spot the problem. In particular the continuum
extrapolations look reasonable even for $c=0.25$. On the other hand
for large enough $c\sim 0.4$, both continuum extrapolations agree
nicely (Fig.~\ref{fig:cont1} (c)). Figs.~\ref{fig:cont1}
(b) and (d) show that the disagreement is larger at strong couplings.

This example illustrates very well the reasons behind the efforts that
different groups have put into understanding cutoff effects of flow
quantities.

\subsection{Tree-level correction}

\newcommand{\basispl}{
   \put(-.5,-.5){\line(1,0){1}}
   \put(.5,-.5){\line(0,1){1}}
   \put(.5,.5){\line(-1,0){1}}
   \put(-.5,.5){\line(0,-1){1}}
                         }
\newcommand{\basisar}{
   \put(0,-.5){\vector(1,0){0}}
   \put(.5,0){\vector(0,1){0}}
   \put(0,.5){\vector(-1,0){0}}
   \put(-.5,0){\vector(0,-1){0}}
	              }
\newcommand{\revar}{
   \put(0,-.5){\vector(-1,0){0}}
   \put(.5,0){\vector(0,-1){0}}
   \put(0,.5){\vector(1,0){0}}
   \put(-.5,0){\vector(0,1){0}}
	              }
\newcommand{\plaq}{\setlength{\unitlength}{.5cm}\raisebox{-.2cm}{
   \begin{picture}(1.2,1.2)(-.6,-.6)
   \basispl\basisar
   \put(-.5,-.5){\circle*{.2}}
   \put(-.55,-.55){\makebox(0,0)[tr]{\footnotesize $x$}}
   \put(-.55,0){\makebox(0,0)[r]{\footnotesize $\nu$}}
   \put(0,-.55){\makebox(0,0)[t]{\footnotesize $\mu$}}
   \end{picture}}}
\newcommand{\twoplaq}{\setlength{\unitlength}{1cm}\raisebox{-.5cm}{
   \begin{picture}(1.2,1.2)(-.6,-.6)
   \basispl
   \put(-.5,-.5){\circle*{.1}}
   \put(-.5,.5){\circle*{.1}}
   \put(.5,-.5){\circle*{.1}}
   \put(.5,.5){\circle*{.1}}
   \put(0,-.5){\circle*{.1}}
   \put(0,.5){\circle*{.1}}
   \put(.5,0){\circle*{.1}}
   \put(-.5,0){\circle*{.1}}
   \put(-.25,-.5){\vector(1,0){0}}
   \put(.25,-.5){\vector(1,0){0}}
   \put(.5,-.25){\vector(0,1){0}}
   \put(.5,.25){\vector(0,1){0}}
   \put(-.25,.5){\vector(-1,0){0}}
   \put(.25,.5){\vector(-1,0){0}}
   \put(-.5,-.25){\vector(0,-1){0}}
   \put(-.5,.25){\vector(0,-1){0}}
   \put(-.55,-.55){\makebox(0,0)[tr]{\footnotesize $x$}}
   \put(-.55,0){\makebox(0,0)[r]{\footnotesize $\nu$}}
   \put(0,-.55){\makebox(0,0)[t]{\footnotesize $\mu$}}
   \end{picture}}}
\newcommand{\stapup}{\setlength{\unitlength}{.5cm}\raisebox{-.2cm}{
   \begin{picture}(1.2,1.2)(-.6,-.6)
   \put(.5,-.5){\line(0,1){1}}
   \put(.5,.5){\line(-1,0){1}}
   \put(-.5,.5){\line(0,-1){1}}
   \put(.5,0){\vector(0,-1){0}}
   \put(0,.5){\vector(1,0){0}}
   \put(-.5,0){\vector(0,1){0}}
   \put(-.5,-.5){\circle*{.2}}
   \put(-.55,-.55){\makebox(0,0)[tr]{\footnotesize $x$}}
   \put(-.55,0){\makebox(0,0)[r]{\footnotesize $\nu$}}
   \put(0,.55){\makebox(0,0)[b]{\footnotesize $\mu$}}
   \end{picture}}}
\newcommand{\stapdw}{\setlength{\unitlength}{.5cm}\raisebox{-.2cm}{
   \begin{picture}(1.2,1.2)(-.6,-.6)
   \put(.5,-.5){\line(0,1){1}}
   \put(.5,-.5){\line(-1,0){1}}
   \put(-.5,.5){\line(0,-1){1}}
   \put(.5,0){\vector(0,1){0}}
   \put(0,-.5){\vector(1,0){0}}
   \put(-.5,0){\vector(0,-1){0}}
   \put(-.5,.5){\circle*{.2}}
   \put(-.55,.75){\makebox(0,0)[tr]{\footnotesize $x$}}
   \put(-.55,0){\makebox(0,0)[r]{\footnotesize $\nu$}}
   \put(0,-.55){\makebox(0,0)[t]{\footnotesize $\mu$}}
   \end{picture}}}
\newcommand{\clover}{\setlength{\unitlength}{.5cm}\raisebox{-.5cm}{
   \begin{picture}(2.4,2.4)(-1.2,-1.2)
   \multiput(-1.2,-1.2)(1.2,1.2){2}{\begin{picture}(1.2,1.2)(-.6,-.6)
   \basispl\basisar\end{picture}}
   \multiput(-1.2,0)(1.2,-1.2){2}{\begin{picture}(1.2,1.2)(-.6,-.6)
   \basispl\revar\end{picture}}
   \put(-.1,-.1){\circle*{.2}}
   \put(-.1,.1){\circle*{.2}}
   \put(.1,-.1){\circle*{.2}}
   \put(.1,.1){\circle*{.2}}
   \end{picture}}}
\newcommand{\twooneplaq}{\setlength{\unitlength}{.5cm}
   \raisebox{-.2cm}{
   \begin{picture}(2.2,1.2)(-1.1,-.6)
   \put(-1,-.5){\line(1,0){2}}
   \put(-1,.5){\line(1,0){2}}
   \put(-1,-.5){\line(0,1){1}}
   \put(1,-.5){\line(0,1){1}}
   \multiput(-1,-.5)(1,0){3}{\circle*{.2}}
   \multiput(-1,.5)(1,0){3}{\circle*{.2}}
   \end{picture}}}
\newcommand{\plaqa}{\setlength{\unitlength}{.5cm}\raisebox{-.2cm}{
   \begin{picture}(1.2,1.2)(-.6,-.6)
   \basispl
   \put(-.5,-.5){\circle*{.2}}
   \put(-.5,.5){\circle*{.2}}
   \put(.5,-.5){\circle*{.2}}
   \put(.5,.5){\circle*{.2}}
   \end{picture}}}
\newcommand{\hookplaq}{\setlength{\unitlength}{.5cm}
   \raisebox{-.3268cm}{
   \begin{picture}(1.7071,1.7071)(-.7071,-.7071)
   \put(0,0){\line(0,1){1}}
   \put(0,1){\line(1,0){1}}
   \put(1,1){\line(0,-1){1}}
   \put(-.7071,-.7071){\line(1,0){1}}
   \put(0,0){\line(-1,-1){.7071}}
   \put(1,0){\line(-1,-1){.7071}}
   \multiput(0,0)(1,0){2}{\circle*{.2}}
   \multiput(0,1)(1,0){2}{\circle*{.2}}
   \multiput(-.7071,-.7071)(1,0){2}{\circle*{.2}}
   \multiput(0,0)(.25,0){4}{\circle*{.03}}
   \end{picture}}}
\newcommand{\cornplaq}{\setlength{\unitlength}{.5cm}
   \raisebox{-.3268cm}{
   \begin{picture}(1.7071,1.7071)(-.7071,-.7071)
   \put(-.7071,-.7071){\line(0,1){1}}
   \put(0,1){\line(1,0){1}}
   \put(1,1){\line(0,-1){1}}
   \put(-.7071,-.7071){\line(1,0){1}}
   \put(0,1){\line(-1,-1){.7071}}
   \put(1,0){\line(-1,-1){.7071}}
   \put(-.7071,-.7071){\circle*{.1}}
   \put(-.7071,.2929){\circle*{.2}}
   \multiput(0,0)(1,0){2}{\circle*{.2}}
   \multiput(0,1)(1,0){2}{\circle*{.2}}
   \multiput(-.7071,-.7071)(1,0){2}{\circle*{.2}}
   \multiput(0,0)(.25,0){4}{\circle*{.03}}
   \multiput(0,0)(0,.25){4}{\circle*{.03}}
   \multiput(0,0)(-.1768,-.1768){4}{\circle*{.03}}
   \end{picture}}}


There are many possible choices for lattice gauge actions, but a
general class of them can be written symbolically as 
\begin{equation}
  \label{eq:act}
  S({c_i}) = \frac{1}{g_0^2}\sum_{x} {\rm Tr} 
      \left( 1 - c_0 \plaqa - c_1 \twooneplaq - 
        c_2 \hookplaq - c_3 \cornplaq\right) \,.
\end{equation}

These include the L\"uscher-Weisz tree-level improved action
($c_0=5/3, c_1=-1/12, c_2=c_3=0$) and the Iwasaki action ($c_0=3.648,
c_1=-0.331, c_2=c_3=0$). 

One has to choose several actions to study a
flow quantity, first the action used for the simulation (i.e. to
produce the configurations), second the action used in the flow
equation~(\ref{eq:flowlat}). Finally, note that 
$\langle E(x,t)\rangle$ is nothing more than an action density, and
therefore, if interested in this observable, any particular
discretization of $\langle E(x,t)\rangle$ can be understood as a third 
choice of action. Each of these actions can be different, and
therefore we have several sets of coefficients $c_i$: one for the
action, one for the flow, and one for the observable.

The influence of these coefficients to tree-level has been clarified 
recently~\cite{Fodor:2014cpa,Ramos:2014kka}. The
perturbative expansion of our observable of interest reads
\begin{equation}
  \langle E(x,t)\rangle = \mathcal E_0(t, a/L) g^2 + \mathcal O(g^4)\,,
\end{equation}
where $\mathcal E_0$ can be computed on the lattice. In fact in the
case of an infinite lattice we have
\begin{equation}
  \label{eq:pt}
  t^2\mathcal E_0(t,a/L) = \frac{3(N^2-1)}{128\pi^2} \left\{
    1 + \frac{a^2}{t} \left[ d^{(o)} - d^{(a)} - 3d^{(f)}\right]
  \right\} + \mathcal O(a^4)\,,
\end{equation}
where the coefficients $d^{(o,a,f)}$ parametrize the tree-level cutoff
effects produced by the observable discretization, the action used in
the simulation and the action used to define the flow respectively. 
The values of these parameters for some popular choices of
discretizations are
\begin{equation}
  d^{(o,a,f)} = \left\{ 
      \begin{array}{l@{\hskip 24pt}l}
        -1/24 & \textrm{plaquette} \\
        1/72  & \textrm{L\"uscher-Weisz} \\
        -5/24 & \textrm{Clover}
      \end{array}
    \right.\,.
\end{equation}

At this point there are
some important observations to make
\begin{enumerate}
\item The particularly popular choice of using the clover
  discretization to 
  define flow quantities (couplings, or the scales $t_0, w_0, \dots$),
  does not 
  seem to be a very good idea from the perspective of perturbation
  theory. This discretization is the one with larger tree-level cutoff
  effects: 5 times larger than the plaquette definition, and 15 times
  larger than the L\"uscher-Weisz discretization. 

\item There is a cancellation between cutoff effects produced by the
  observable and those produced by the action and the flow. For
  example the choice of a Wilson action, Wilson flow, and clover
  observable leads to rather small tree-level cutoff effects. This is
  due to an accidental cancellation of the cutoff effects produced by
  the action and flow on one hand, and those of produced by the
  observable on the other. Note that this accidental cancellation
  disappears as soon as the setup is changed. For example, using the
  Symanzik flow and L\"uscher-Weisz action, but keeping the clover
  observable leads to a very poor scenario, with tree-level cutoff
  effects  around 7 times larger.  
\end{enumerate}

The perturbative computation of Eq.~(\ref{eq:pt}) immediately suggests
an improved definition for the coupling
\begin{equation}
  \label{eq:gimp}
  g_{\rm GF}^2(L) = \hat{\mathcal N}^{-1}t^2\langle
  E(x,t)\rangle\Big|_{\sqrt{8t} = cL}\,,
\end{equation}
where $\hat{\mathcal N} = \mathcal E_0$ (i.e. Using the tree-level
lattice computation instead of the continuum result to define the
coupling). This definition 
of the coupling has no cutoff effects to tree-level, 
and has been used in most step scaling studies
(see~\cite{Fritzsch:2013je,Ramos:2014kla,Luscher:2014kea,Lin:2014fxa,Perez:2014isa}). The
exception were the studies using periodic boundary conditions 
(like~\cite{Fodor:2012td,Cheng:2014jba,Hasenfratz:2014rna}), or
smeared gauge actions (for example~\cite{Rantaharju:2013bva}), since the 
perturbative computation in these setups is more involved (see the
recent work~\cite{Fodor:2014cpa} for the case of periodic boundary
conditions).  

\begin{figure}
  \centering
  \includegraphics[width=0.6\textwidth]{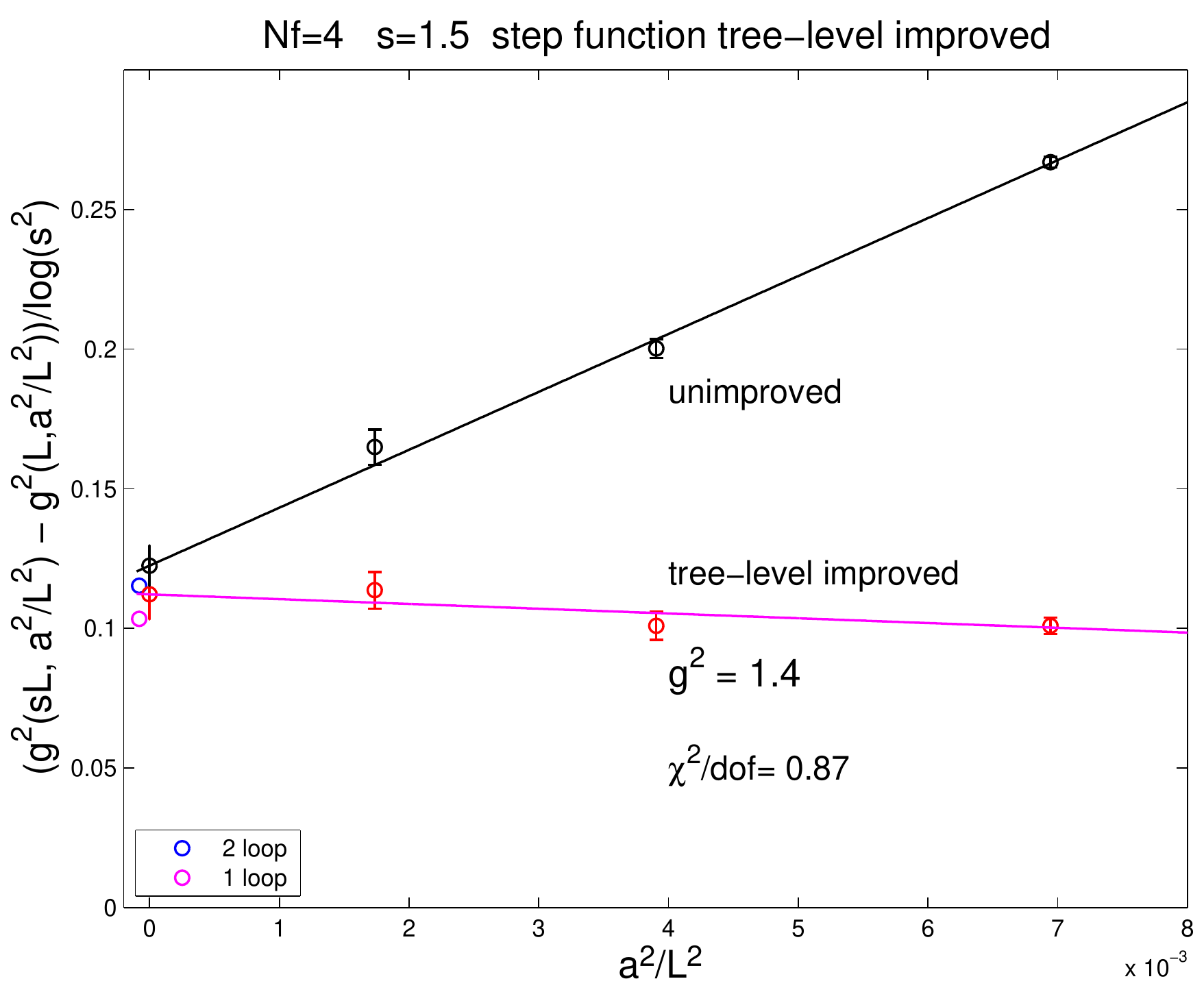}
  \caption{Comparison of the continuum extrapolation of the step
    scaling function of the coupling defined with the continuum norm
    and the tree-level improved
    definition. (Source~\cite{Fodor:2014cpa}). }
  \label{fig:tlimp}
\end{figure}

This improved definition of the coupling has a big impact on the
cutoff effects. A nice example is given in~\cite{Fodor:2014cpa}, where
the step-scaling study of~\cite{Fodor:2012td} is repeated with the
improved definition of the coupling Eq.~(\ref{eq:gimp}). The result
can be seen in Fig.~\ref{fig:tlimp}. This example is specially
impressive, partially because the original data of~\cite{Fodor:2012td}
has a choice of discretizations (L\"uscher-Weisz action, Symanzik
flow and Clover observable) that, as we have commented, make the
tree-level correction specially large. 

The work~\cite{Fodor:2012td} also proposes to use a general class
of actions given by Eq.~(\ref{eq:act}) setting $c_2=c_3=0$. Since the
two remaining coefficients are related by $c_0 + 8c_1=1$, each choice
of action has one free parameter, resulting in three parameters that
can be chosen arbitrarily. The concrete proposal is to choose them to
cancel the tree-level cutoff effects of $\langle E(x,t)\rangle$ in
infinite volume to order $a^2, a^4$ and $a^6$. It
is the opinion of the author that this criteria to choose the
coefficients is not very appealing, especially taking into account that
one can remove all the tree-level cutoff effects (to all orders in
$a$) with the simple coupling definition of Eq.(\ref{eq:gimp}), and
there is no good reason to think that this choice of coefficients will
result in reduced cutoff effects for any other quantity or to higher
orders in $g_0$.

\subsection{Phenomenological improvement}

In any $\mathcal O(a)$ improved setup, the lattice gradient flow
coupling can be expanded as  
\begin{equation}
  g_{\rm GF}^2(L; a/L) = \mathcal N^{-1}t^2\langle
  E(x,t)\rangle\Big|_{\sqrt{8t} = cL} = 
  g_{\rm GF}^2(L; 0) + \left(\frac{a}{L}\right)^2 A(c,L) + \mathcal O(a^4)\,,
\end{equation}
where, as we have seen in the previous section, $A(c,L)$ depends not
only on the volume and the parameter $c$, but also on the choice of
boundary conditions, the action, the flow, etc\dots The proposal
of~\cite{Cheng:2014jba} is to subtract the cutoff effects  $a^2A(c,L)$
by modifying the definition of the coupling to 
\begin{equation}
  \tilde g_{\rm GF}^2(L; a/L) = \mathcal N^{-1}t^2\langle
  E(x,t+a^2\tau)\rangle\Big|_{\sqrt{8t} = cL}\,.
\end{equation}
The parameter $\tau$ is chosen to cancel the leading cutoff
effects. One can see this $t$-shift as adding an $a^2$ term 
to the coupling definition
\begin{equation}
  \tilde g_{\rm GF}^2(L; a/L) = g_{\rm GF}^2(L; a/L)
  + \frac{a^2\tau}{\mathcal N}\langle
  t^2\partial_tE(x,t)\rangle\Big|_{\sqrt{8t} = cL} + \mathcal O(a^4)\,.
\end{equation}

Recently a similar approach in spirit has been taken
in~\cite{Fodor:2015baa}. In this case
instead of canceling the $\mathcal O(a^2)$ term by a shift in
the flow time, the authors use a linear combination of the coupling
defined with different discretizations of the observable
\begin{equation}
  \tilde g_{\rm GF}^2(L; a/L) = X g_{\rm GF; Clover}^2(L; a/L) +
  (1-X) g_{\rm GF; Plaquette}^2(L; a/L)\,,
\end{equation}
where the parameter $X$ is chosen to eliminate the leading cutoff effects. 

\begin{figure}
  \centering
  \subfloat[][]{
    \includegraphics[width=0.45\textwidth]{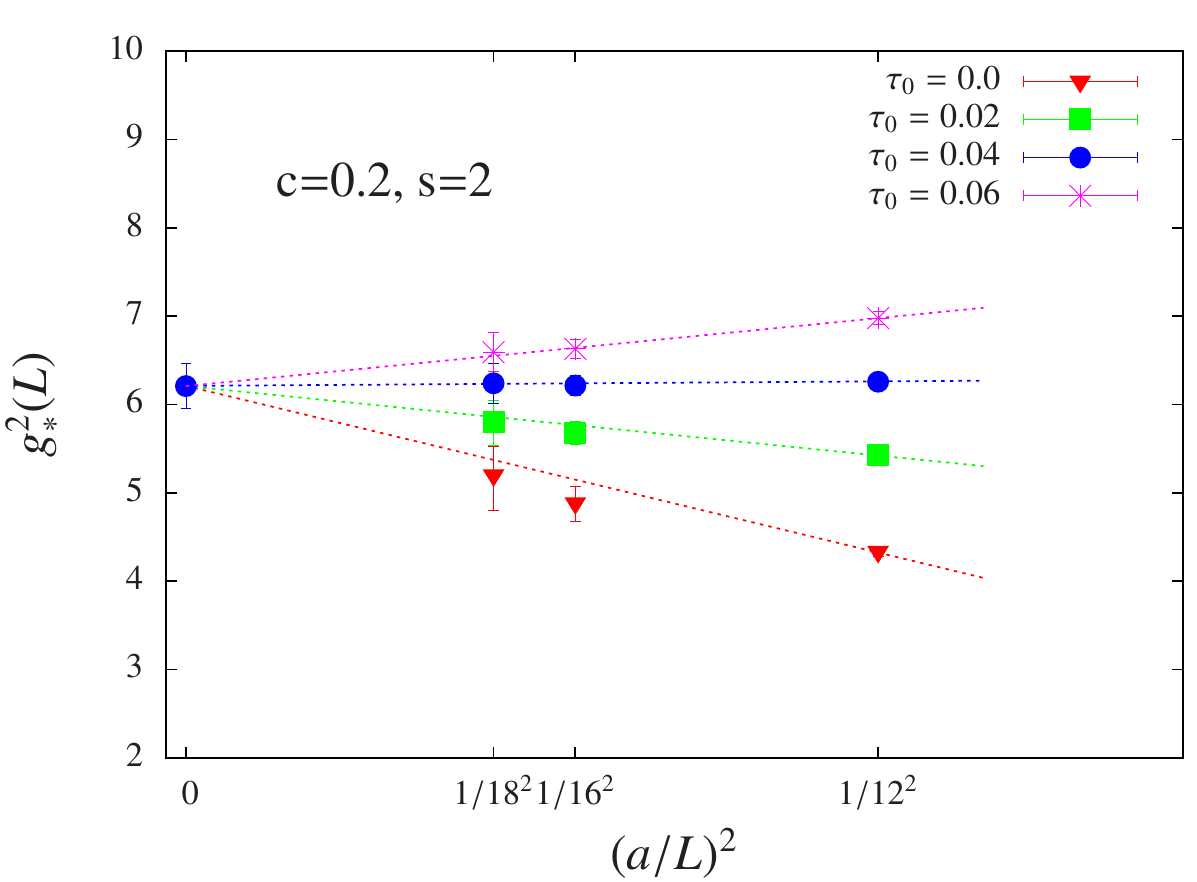}
  }
  \subfloat[][]{
    \includegraphics[width=0.45\textwidth]{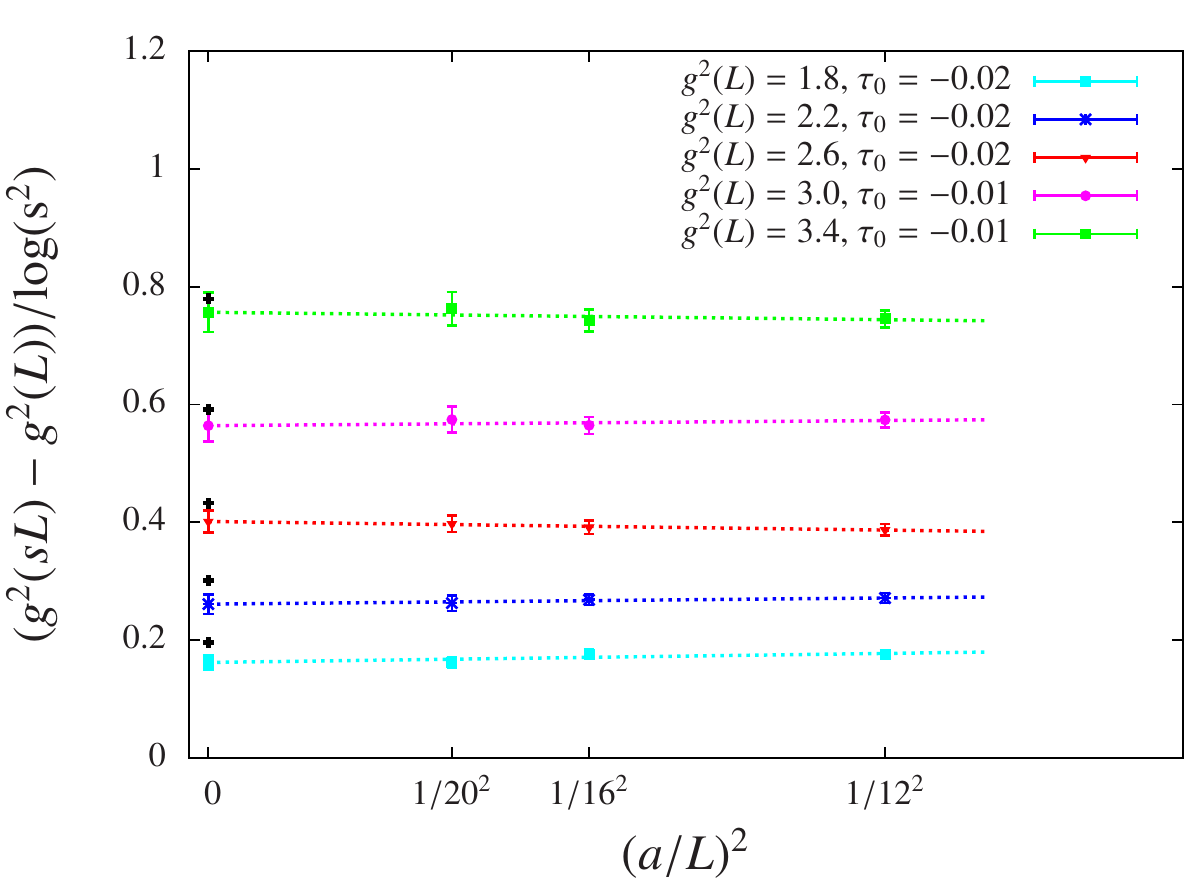}
  } 
  \caption{Left: Continuum extrapolation of the step-scaling function
    for different values of $\tau$}
  \label{fig:tau}
\end{figure}

Obviously these are always legitimate things to do,
that cannot change the values obtained after a properly done continuum
extrapolation. In this context one should avoid giving a dependence to
$\tau$ or $X$ 
on the bare parameters (like $\beta$), since this would result in a
different $a^2$ shift for each value of $a$, and could give the wrong
continuum extrapolations. One should also be careful with
extrapolating to the continuum with a constant fit, even if the data
``looks flat'' after having
applied the correction, since one could artificially
reduce the statistical error of the extrapolation. 
Fig.~\ref{fig:tau} shows how different values of $\tau$
results in data with different cutoff effects, but whose
continuum extrapolations are in agreement. Moreover these recipes are
simple to implement since the integration of the flow equations
naturally gives $\langle E(x,t)\rangle$ at many values of $t$, and the
evaluation of a different observable is always cheap. 

In the opinion of the author the weak point of these approaches is that
the improvement of flow quantities requires one to improve the flow
equation, the discretization of the observable \emph{and} the action
used to generate the
configurations~\cite{Sommer:2014mea,Ramos:2014kka}. A single parameter
($\tau$ or $X$) is not able to achieve this effect and this translates in
$\tau$ (or $X$) being a function of everything: choice of boundary
conditions, 
choice of action (both gauge and fermionic), volume, etc\dots The
dependence of this parameter on these details is not small (see
for example Fig.~\ref{fig:tau}, where $\tau$ changes by 100\% when
the renormalized coupling changes in the range $\sim 2-4$). In the end
$\tau$ (or $X$) cannot be computed analytically nor by
studying other flow quantities, but has to 
be inferred from the same data that one is trying to improve. 

\subsection{Symanzik improvement and the Zeuthen flow}

A very appealing approach to systematically reduce cutoff effects is
the Symanzik improvement program. Its purpose is far more ambitious
than looking for an accidental cancellation between cutoff effects for some
particular observable. When we choose a particular discretization (to
define an action/observable) we are choosing what our theory looks
like at very high energy scales (the cutoff $1/a$). The idea of the
Symanzik improvement program is to choose these discretizations so
that the effective theory at energy scales much smaller than the
cutoff looks as close as possible to the continuum theory. This
program starts by describing a particular lattice action $S_{\rm latt}$
by an effective continuum theory. For example for pure gauge theories
any lattice action can be described by an effective action given by
\begin{equation}
  S_{\rm eff} = S_{\rm YM} + a^2S_2 + a^4 S_4 + \dots\,,
\end{equation}
(i.e. any choice of actions will have the Yang-Mills action as first
approximation $S_{\rm YM}$, but different discretizations will produce
different cutoff effects described by $S_2, S_4, \dots$). Symanzik's
idea was to choose a discretization such that the effective continuum
description has some of these terms absent (i.e. $S_2=0$). 

We can say that the basic idea behind Symanzik improvement program is
to help us choosing a discretization that represents better continuum
physics. Being a choice made at the cutoff scale $1/a$, \emph{it should
  be obvious that improvement parameters can not depend\footnote{Up to small
    logarithmic corrections.} on the boundary
  conditions, the volume, or the choice of observable\footnote{One should
    nevertheless use an improved definition for the observable if one
    wants cutoff effects be reduced for a particular quantity.}}. In
other words, if one finds the Symanzik improved action, this is
valid for all observables (this includes all values of the parameter
$c$ that defines the coupling), all volumes, all boundary
conditions,etc\dots 

The application of the Symanzik improvement program to the gradient
flow has to be done using a 5D local field
theory~\cite{Luscher:2013cpa}, in which the flow time becomes the 
fifth coordinate with values in the positive real
axis. The action in the continuum Euclidean space can be written as
\begin{equation}
      S^{\rm cont} = -\frac{1}{2g_0^2} \int{\rm d}^4x\,  
      {\rm Tr}\left\{
        F_{\mu\nu}F_{\mu\nu}
      \right\} -
      2\int_0^\infty {\rm d} t\int{\rm d}^4x\,
      {\rm Tr}\left\{ { L_\mu(x,t)}[\partial_t B_{\mu}(x,t) - 
          D_\nu G_{\nu\mu}]
      \right\}\,,
\end{equation}
where $F_{\mu\nu}$ and $G_{\mu\nu}$ are the field strength at $t=0$
and $t>0$ respectively. $L_\mu(x,t)$ is a Lagrange
multiplier in the sense that the path integral over
this field imposes the flow equation~(\ref{eq:flow}) in the bulk of the
5D space (i.e. for $t>0$). 

The main point is that due to the classical nature of the theory at
$t>0$, one can completely eliminate all (i.e. to
all orders in the coupling) $\mathcal O(a^2)$ cutoff
effects in the bulk~\cite{Ramos:2014kka}. Making a long story short,
this non-perturbative improvement is achieved by using
the \emph{Zeuthen flow} to integrate the flow equations
\begin{equation}
  a^2\partial_t V_\mu(x,t) = -g_0^2 \left[
    \left(1 + \frac{a^2}{12}D_\mu D_\mu^* \right) T^a\partial_{x,\mu}^a
    S_{\rm LW}(V)\right]  
  V_\mu(x,t)\,,
\end{equation}
where $D_\mu$ and $D_\mu^*$ are the forward and backward lattice
covariant derivatives, and $S_{\rm LW}$ is the tree-level
L\"uscher-Weisz improved action (i.e. Eq.(\ref{eq:act}) with $c_0=5/3,
c_1=-1/12, c_2=c_3=0$). In the same way one can define clasically
improved discretizations for observables (like $E(x,t)$), that will
not introduce any $\mathcal O(a^2)$ discretization effects. 

In summary, from the three discretization choices that one has to make
in order to study any flow quantity, the Symanzik improvement program
gives a way to fix the discretization used to integrate the flow
equation and to evaluate flow observables, so that no $\mathcal
O(a^2)$ effects are introduced. The remaining cutoff effects are only
those of 
the action used to produce the configurations and a couple of
boundary counterterms. 

\section{Some recent results}

Since the proposal of using the gradient flow to
define a renormalized coupling there have been several works
using this technique to investigate the dynamics of strongly coupled
non-abelian gauge theories. Most of the applications are related to
models beyond QCD and the search for an infrared fixed-point (see for a
recent review~\cite{Aoki:2014latt}). On the 
other hand, the gradient flow has been used in QCD mainly for scale
setting (see~\cite{Sommer:2014mea}). We have already mentioned some of
the recent works in the previous sections. In this section we will
briefly mention other works paying attention to the details discussed in
earlier sections. The reader interested in a more exhaustive review of
the recent works should also consult~\cite{Aoki:2014latt}. 

\begin{figure}
  \centering
  \subfloat[][]{
    \includegraphics[width=0.45\textwidth]{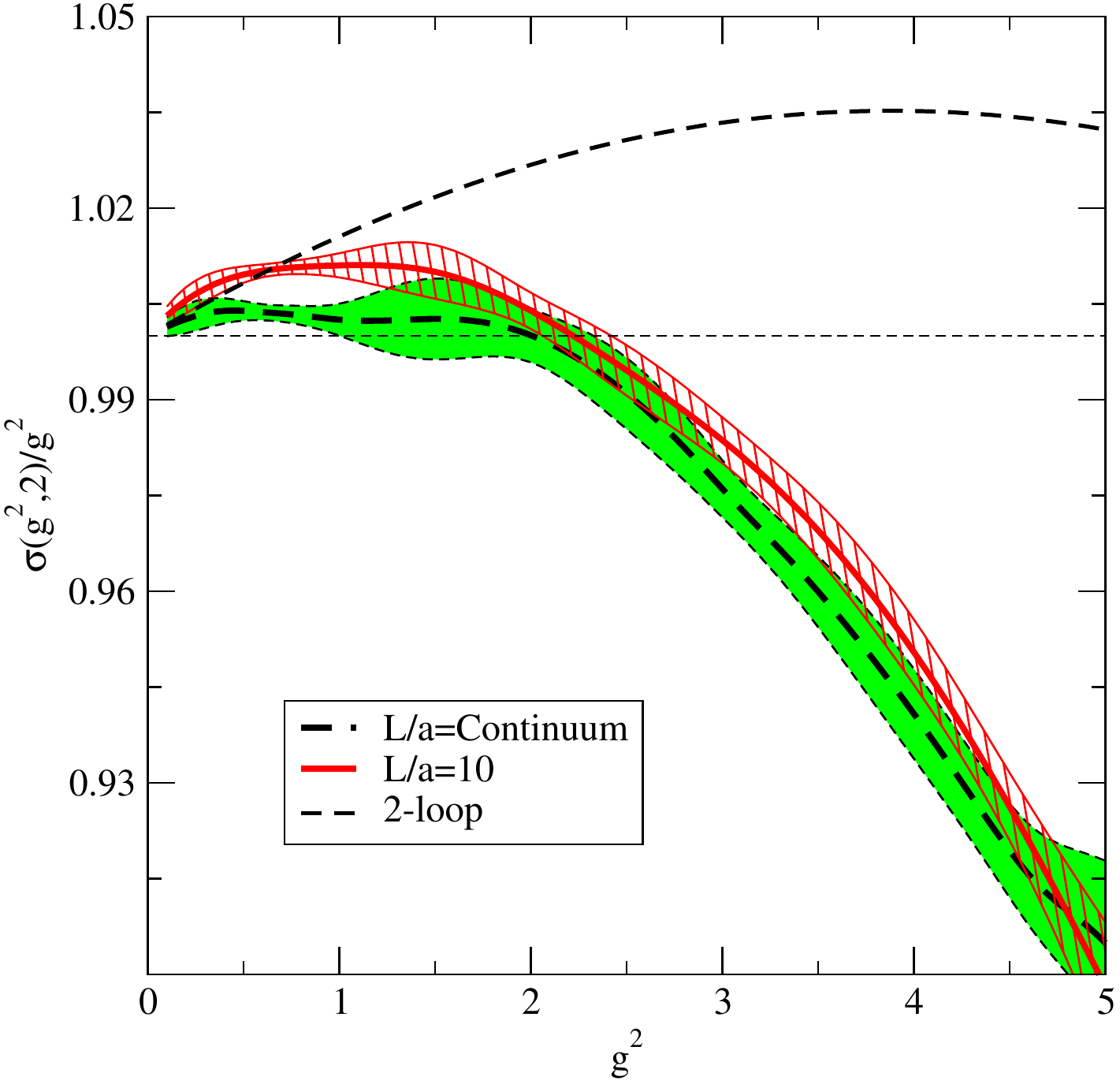}
  }
  \subfloat[][]{
    \includegraphics[width=0.45\textwidth]{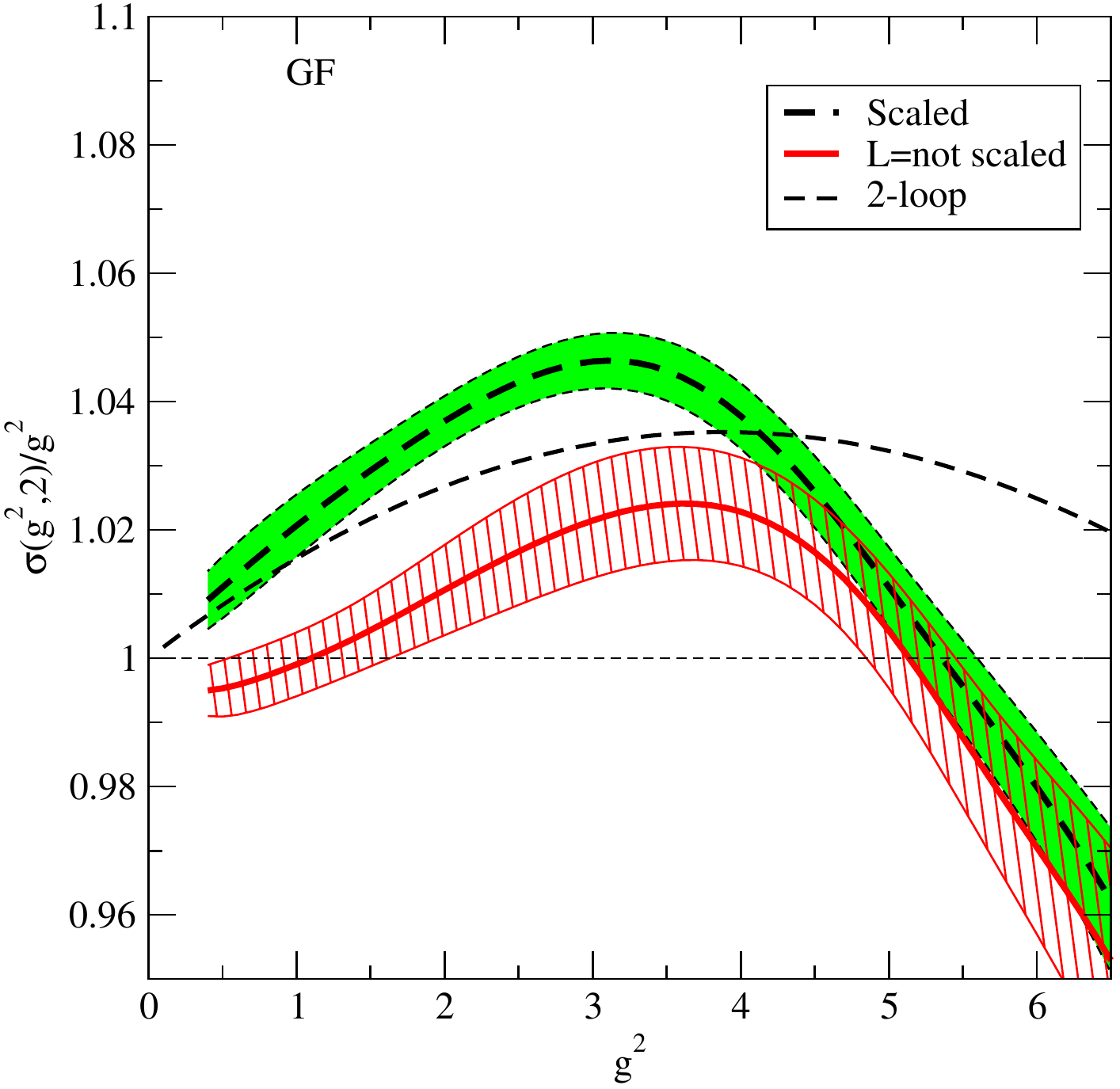}
  } 
  \caption{Step scaling function in the minimal walking
    technicolor in two different schemes. Left:  $\sigma_2(g^2_{\rm
      SF})/g_{\rm SF}^2$ in the SF scheme, showing an IR fixed point
    at $g^2_{\rm SF}\sim 2$. Right: $\sigma_2(g^2_{\rm GF})/g_{\rm GF}^2$ in
    the gradient flow scheme, showing an IR fixed point at $g^2_{\rm
      GF}\sim 5$. (Source~\cite{Rantaharju:2013bva})} 
  \label{fig:su2ad}
\end{figure}

The work~\cite{Rantaharju:2013bva} studies the 
running of the coupling in the minimal technicolor model ($SU(2)$ gauge
group with two adjoint fermions) using both the Schr\"odinger
functional coupling and the gradient flow coupling. Both schemes show
an IR fixed 
point (see Fig~\ref{fig:su2ad}) at quite different values of the
coupling. Of course since the schemes are different there is no reason
to expect the values of the coupling at the fixed point to be the
same. This work uses a smeared gauge action,
which modifies the lattice leading order value of $t^2\langle
E(x,t)\rangle$.  
The improved coupling definition~Eq.(\ref{eq:gimp}) cannot be used
without an analytic computation. The authors address the issue of
large cutoff effects by using a
large value of $c$

\begin{figure}
  \centering
  \subfloat[][]{
    \includegraphics[width=0.4\textwidth,angle=-90]{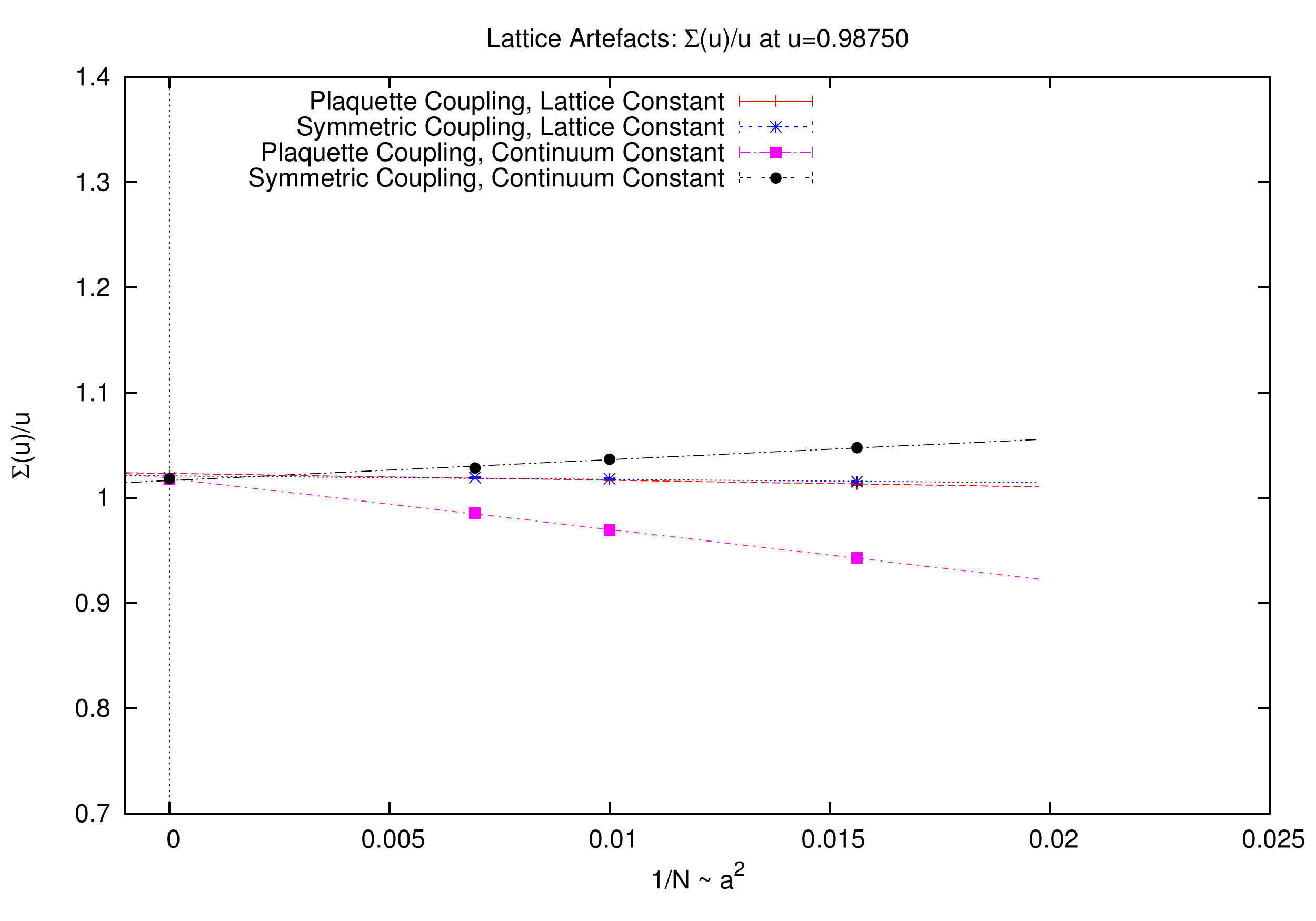}
  }
  \subfloat[][]{
    \includegraphics[width=0.4\textwidth,angle=-90]{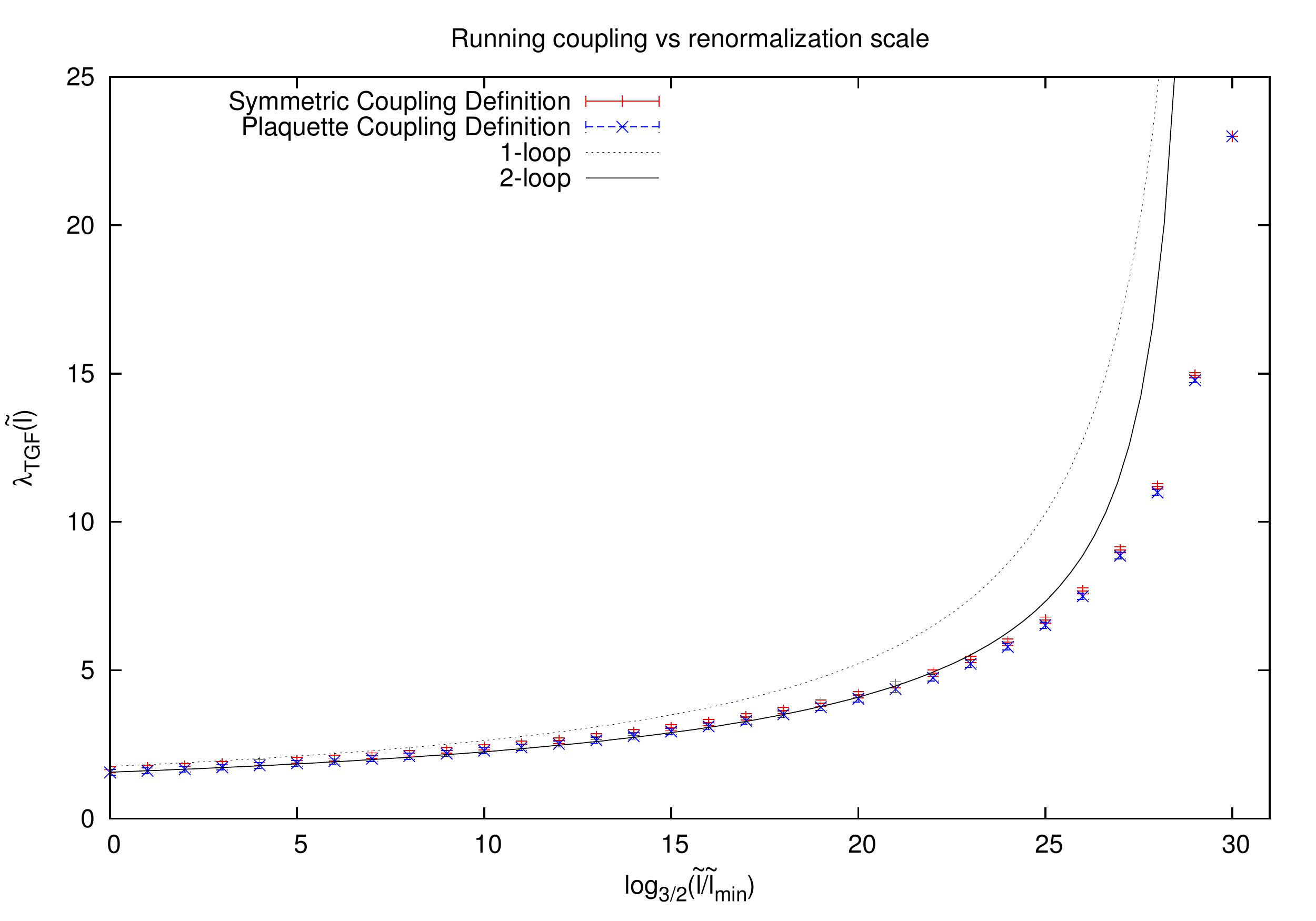}
  } 
  \caption{Left: Continuum extrapolation of the step scaling function
    of the t'Hooft coupling in $SU(\infty)$. Right: Running of the
    t'Hooft coupling $\lambda = Ng^2$ in $SU(\infty)$ gauge theory
    (source~\cite{Perez:2014isa}).} 
  \label{fig:tek}
\end{figure}

Computing non-perturbative effects in $SU(N)$ gauge theories in the
large $N$ 't Hooft limit (keeping $\lambda = g^2N$ constant) via
lattice techniques is a difficult task because 
the number of degrees of freedom that have to be simulated on a
computer increase with $N$. Nevertheless, using the ideas of volume
independence (see~\cite{Perez:2014sqa} for a recent review), the
authors of~\cite{Perez:2014isa} have computed the running
of the t'Hooft coupling Fig.~\ref{fig:tek}. 
The authors use the tree-level corrected coupling definition
Eq.~(\ref{eq:gimp}) and the clover and plaquette definitions of the
observable to estimate the systematic of the continuum
extrapolations. Their results also indicate that the tree-level
corrected definition of the coupling 
Eq.~(\ref{eq:gimp}) has a big impact in the cutoff effects. 

\section{Fermion flow and renormalization of composite operators} 

The flow can also be introduced for the fermion fields~\cite{Luscher:2013cpa}
\begin{equation}
  \partial_t \chi(x,t) = D_\mu D_\mu \chi(x,t)\,,
\end{equation}
where $D_\mu = \partial_\mu + B_\mu$, and the fermion flow field obeys
the initial condition $\chi(x,t)|_{t=0} = \psi(x)$. In this case the
fermion flow field requires renormalization, but there is no operator
mixing at positive flow time. i.e. composite fermion operators
renormalize multiplicatively at positive flow time. 

In fact the study of Ward Identities at positive flow time has lead to
a novel approach to the renormalization of some composite
operators. We have recently seen the cases of the axial current, the
chiral condensate~\cite{Luscher:2013cpa,Shindler2014}, or the energy
momentum tensor~\cite{DelDebbio:2013zaa}. 

On the other hand the so called small flow-time expansion offers the
possibility of 
a general approach to the renormalization of a generic composite
operator $O(x,t)$. When the flow time is taken to zero $O(x,t)$ has to
converge to the bare operator $O(x)$. Since the
latter usually requires renormalization, $O(x,t)$ becomes singular 
when taking the $t\rightarrow 0$ limit in a way that can be expressed as 
an asymptotic expansion in renormalized operators with singular
coefficients~\cite{Luscher:2011bx}
\begin{equation}
  O(x,t) = \sum_\alpha c_\alpha(t) O_{\rm R}^\alpha(x) + \mathcal O(t)\,.
\end{equation}
The previous sum runs over all the operators of dimension smaller that
the dimension of $O(x,t)$ and compatible with its
\emph{continuum} symmetries. We can use $E(x,t) = -\frac{1}{4}{\rm
  Tr}(G_{\mu\nu}G_{\mu\nu})$, as an example, whose small flow time
expansion reads 
\begin{equation}
  E(x,t) = c_1(t)\mathbf 1 + c_2(t)\{F_{\mu\nu}F_{\mu\nu}\}_{\rm R}(x)
  + \mathcal O(t)\,.
\end{equation}
In this example, $E(x,t)$ mixes with the vacuum and with the
renormalized action density. In fact this relation 
can be used to determine the spin-0 component of the energy-momentum
tensor~\cite{Suzuki:2013gza}
\begin{equation}
  \{T_{\mu\mu}\}_{\rm R}(x) = \{F_{\mu\mu}F_{\mu\mu}\}_{\rm R}(x) = 
  \lim_{t\rightarrow 0} c_2^{-1}(t)\left\{
    E(x,t) - \langle E(x,t)\rangle
  \right\}\,.
\end{equation}

This relation holds in the continuum, and its use in the lattice
framework \emph{requires that the continuum limit is taken before the}
  $t\rightarrow 0$ \emph{limit}. This means that the scale given by
the flow time has to be large compared with the lattice spacing, and
small compared with any other relevant scale of the problem
(i.e. $a\ll\sqrt{8t}\ll L,\Lambda,T$). 

Moreover one needs to compute the small flow time coefficients
$c_\alpha(t)$. Ref.~\cite{Suzuki:2013gza} proposes to use perturbation
theory. This proposal has recently been used to extract thermodynamic
quantities (see the contribution of this
conference~\cite{Kitazawa:2014uxa} for more details). Generally
speaking there might be concerns about using leading order
perturbation theory at energy scales below the lattice cutoff $1/a$,
and this is the reason why we have seen different works focused on the
non-perturbative determination of the small flow time
coefficients. The authors of~\cite{patellaconf} propose to determine
non-perturbatively the quantity $\gamma_\alpha(t) = -2t \frac{d}{dt}\log
  c_\alpha(t)$ and then construct $c_\alpha(t)$ by integrating
using perturbation theory as initial condition. The functions
$\gamma_\alpha(t)$ can be extracted from correlation functions. For
example in the case mentioned before we have 
\begin{equation}
  \gamma_2(t) = -2t\frac{d}{dt} \log \langle E(x,t)E(x,s)\rangle_{\rm connected} +
  \mathcal O(t)\,.
\end{equation}

Although the idea is appealing, and has the advantage that it
can be generally applicable to any composite operator with a
complicated mixing pattern (including fermion composite operators), it
remains to be seen how well it actually works in a concrete numerical
example.   

We have also seen how the small flow-time expansion can be used to
compute the strange content of the
nucleon~\cite{Shindler:2014oha}. In this case we consider the
connected part of the scalar strange density $\overline s s(x,t)$ in a
nucleon state 
\begin{equation}
  O^{\rm sub}(x,t) = N\,\overline s s(x,t)\, N^\dagger - 
  \langle N N^\dagger\rangle \langle \overline s s(x,t) \rangle\,,
\end{equation}
where $N$ is a nucleon interpolating operator. The small flow time
expansion of this operator reads
\begin{equation}
  O^{\rm sub}(x,t) = c_3(t)O^{\rm sub}(x,0) + \mathcal O(t)\,,
\end{equation}
and $c_3(t)$ can be determined from the Ward identities
(see~\cite{Shindler:2014oha} for more details).

\section{Conclusions}

We have seen in the recent years a growing interest in the 
gradient flow as a tool for scale setting and to define renormalized
couplings. In this contribution we have reviewed the main ideas behind
these applications putting a strong emphasis on the role of the
continuum extrapolation and cutoff effects of flow quantities. We have
also commented on the use of the small flow-time expansion as a
general tool to renormalize composite operators, a very interesting
and promising application where there is still a lot of work to do. 

I would like to conclude with a brief comment on some recent works
that I did not have time to 
cover or did not fit in the presentation, but nevertheless I found
interesting. Perturbative computations have proven to be a useful tool
(i.e. to understand the structure of cutoff effects), but also
perturbative computations of flow quantities are usually tedious. The
application of numerical stochastic perturbation
theory for flow quantities~\cite{Brida:2013mva} is 
an interesting idea. The use of the gradient flow to extract matrix
elements non-perturbatively with a simplified mixing pattern has also
been considered recently~\cite{Monahan:2015lha}, and shows a promising
future. 

\section*{Acknowledgments}

I want to thank M. Dalla Brida, P. Fritzsch,
M. Garcia Perez, A. Gonzalez-Arroyo, P. Korcyl, D. Lin, S. Schafer, 
H. Simma, S. Sint, R. Sommer, U. Wolff for their help in preparing
this talk and the proceedings. I would like to show my gratitude to
L. Del Debbio, L. Keegan and S. Sint for their help preparing this
manuscript.  

An important part of preparing this talk for the Lattice symposium
consisted in understanding the ideas behind the many works
that I have tried to cover. The help and patience of A. Hasenfratz,
L. Keegan, A. Patella, D. Nogradi, J. Rantaharju, C. Monahan and 
A. Shindler answering my questions has been invaluable. 

\bibliography{/home/alberto/docs/bib/math,/home/alberto/docs/bib/campos,/home/alberto/docs/bib/fisica,/home/alberto/docs/bib/computing}

\providecommand{\href}[2]{#2}\begingroup\raggedright\begin{thebibliography}{10}

\bibitem{Narayanan:2006rf}
R.~Narayanan and H.~Neuberger, {\it {Infinite N phase transitions in continuum
  Wilson loop operators}},  {\em JHEP} {\bf 0603} (2006) 064,
  [\href{http://xxx.lanl.gov/abs/hep-th/0601210}{{\tt hep-th/0601210}}].

\bibitem{Luscher:2009eq}
M.~L{\"u}scher, {\it {Trivializing maps, the Wilson flow and the HMC
  algorithm}},  {\em Commun.Math.Phys.} {\bf 293} (2010) 899--919,
  [\href{http://xxx.lanl.gov/abs/0907.5491}{{\tt arXiv:0907.5491}}].

\bibitem{Lohmayer:2011si}
R.~Lohmayer and H.~Neuberger, {\it {Continuous smearing of Wilson Loops}},
  {\em PoS} {\bf LATTICE2011} (2011) 249,
  [\href{http://xxx.lanl.gov/abs/1110.3522}{{\tt arXiv:1110.3522}}].

\bibitem{Luscher:2010iy}
M.~L{\"u}scher, {\it {Properties and uses of the Wilson flow in lattice QCD}},
  {\em JHEP} {\bf 1008} (2010) 071,
  [\href{http://xxx.lanl.gov/abs/1006.4518}{{\tt arXiv:1006.4518}}].

\bibitem{Luscher:2011bx}
M.~L{\"u}scher and P.~Weisz, {\it {Perturbative analysis of the gradient flow
  in non-abelian gauge theories}},  {\em JHEP} {\bf 1102} (2011) 051,
  [\href{http://xxx.lanl.gov/abs/1101.0963}{{\tt arXiv:1101.0963}}].

\bibitem{Luscher:2013cpa}
M.~Luscher, {\it {Chiral symmetry and the Yang--Mills gradient flow}},  {\em
  JHEP} {\bf 1304} (2013) 123, [\href{http://xxx.lanl.gov/abs/1302.5246}{{\tt
  arXiv:1302.5246}}].

\bibitem{Kitazawa:2014uxa}
M.~Kitazawa, M.~Asakawa, T.~Hatsuda, T.~Iritani, E.~Itou, et~al., {\it
  {Measurement of thermodynamics using gradient flow}},
  \href{http://xxx.lanl.gov/abs/1412.4508}{{\tt arXiv:1412.4508}}.

\bibitem{Sommer:2014mea}
R.~Sommer, {\it {Scale setting in lattice QCD}},  {\em PoS} {\bf LATTICE2013}
  (2014) 015, [\href{http://xxx.lanl.gov/abs/1401.3270}{{\tt
  arXiv:1401.3270}}].

\bibitem{Luscher:1991wu}
M.~L{\"u}scher, P.~Weisz, and U.~Wolff, {\it {A Numerical method to compute the
  running coupling in asymptotically free theories}},  {\em Nucl.Phys.} {\bf
  B359} (1991) 221--243.

\bibitem{Fodor:2012td}
Z.~Fodor, K.~Holland, J.~Kuti, D.~Nogradi, and C.~H. Wong, {\it {The Yang-Mills
  gradient flow in finite volume}},  {\em JHEP} {\bf 1211} (2012) 007,
  [\href{http://xxx.lanl.gov/abs/1208.1051}{{\tt arXiv:1208.1051}}].

\bibitem{Fritzsch:2013je}
P.~Fritzsch and A.~Ramos, {\it {The gradient flow coupling in the Schr\"odinger
  Functional}},  {\em JHEP} {\bf 1310} (2013) 008,
  [\href{http://xxx.lanl.gov/abs/1301.4388}{{\tt arXiv:1301.4388}}].

\bibitem{Ramos:2014kla}
A.~Ramos, {\it {The gradient flow running coupling with twisted boundary
  conditions}},  {\em JHEP} {\bf 1411} (2014) 101,
  [\href{http://xxx.lanl.gov/abs/1409.1445}{{\tt arXiv:1409.1445}}].

\bibitem{Luscher:2014kea}
M.~L\"uscher, {\it {Step scaling and the Yang-Mills gradient flow}},  {\em
  JHEP} {\bf 1406} (2014) 105, [\href{http://xxx.lanl.gov/abs/1404.5930}{{\tt
  arXiv:1404.5930}}].

\bibitem{Symanzik:1981wd}
K.~Symanzik, {\it {Schrodinger Representation and Casimir Effect in
  Renormalizable Quantum Field Theory}},  {\em Nucl.Phys.} {\bf B190} (1981) 1.

\bibitem{Luscher:1992an}
M.~L{\"u}scher, R.~Narayanan, P.~Weisz, and U.~Wolff, {\it {The Schr{\"o}dinger
  Functional: a renormalizable probe for non-abelian gauge theories}},  {\em
  Nucl.Phys.} {\bf B384} (1992) 168--228,
  [\href{http://xxx.lanl.gov/abs/hep-lat/9207009}{{\tt hep-lat/9207009}}].

\bibitem{Sint:1993un}
S.~Sint, {\it {On the Schr{\"o}dinger functional in QCD}},  {\em Nucl.Phys.}
  {\bf B421} (1994) 135--158,
  [\href{http://xxx.lanl.gov/abs/hep-lat/9312079}{{\tt hep-lat/9312079}}].

\bibitem{Sint:2010eh}
S.~Sint, {\it {The Chirally rotated Schr\"odinger functional with Wilson
  fermions and automatic O(a) improvement}},  {\em Nucl.Phys.} {\bf B847}
  (2011) 491--531, [\href{http://xxx.lanl.gov/abs/1008.4857}{{\tt
  arXiv:1008.4857}}].

\bibitem{Frezzotti:2003ni}
R.~Frezzotti and G.~Rossi, {\it {Chirally improving Wilson fermions. 1. O(a)
  improvement}},  {\em JHEP} {\bf 0408} (2004) 007,
  [\href{http://xxx.lanl.gov/abs/hep-lat/0306014}{{\tt hep-lat/0306014}}].

\bibitem{GonzalezArroyo:1981vw}
A.~Gonzalez-Arroyo, J.~Jurkiewicz, and C.~Korthals-Altes, {\it {Ground state
  metamorphosis for Yang-Mills fields on a finite periodic lattice}},  {\em
  Freiburg ASI 1981:0339} (1981).

\bibitem{DelDebbio:2004xh}
L.~Del~Debbio, G.~M. Manca, and E.~Vicari, {\it {Critical slowing down of
  topological modes}},  {\em Phys.Lett.} {\bf B594} (2004) 315--323,
  [\href{http://xxx.lanl.gov/abs/hep-lat/0403001}{{\tt hep-lat/0403001}}].

\bibitem{Schaefer:2010hu}
{\bf ALPHA} Collaboration, S.~Schaefer, R.~Sommer, and F.~Virotta, {\it
  {Critical slowing down and error analysis in lattice QCD simulations}},  {\em
  Nucl. Phys.} {\bf B845} (2011) 93--119,
  [\href{http://xxx.lanl.gov/abs/1009.5228}{{\tt arXiv:1009.5228}}].

\bibitem{Fritzsch:2013yxa}
P.~Fritzsch, A.~Ramos, and F.~Stollenwerk, {\it {Critical slowing down and the
  gradient flow coupling in the Schr\"odinger functional}},  {\em PoS} {\bf
  Lattice2013} (2013) 461, [\href{http://xxx.lanl.gov/abs/1311.7304}{{\tt
  arXiv:1311.7304}}].

\bibitem{Fritzsch:2013hda}
P.~Fritzsch and A.~Ramos, {\it {Studying the gradient flow coupling in the
  Schrödinger functional}},  {\em PoS} {\bf Lattice2013} (2014) 319,
  [\href{http://xxx.lanl.gov/abs/1308.4559}{{\tt arXiv:1308.4559}}].

\bibitem{deDivitiis:1994yz}
{\bf ALPHA} Collaboration, G.~de~Divitiis et~al., {\it {Universality and the
  approach to the continuum limit in lattice gauge theory}},  {\em Nucl.Phys.}
  {\bf B437} (1995) 447--470,
  [\href{http://xxx.lanl.gov/abs/hep-lat/9411017}{{\tt hep-lat/9411017}}].

\bibitem{Brida:2014joa}
M.~D. Brida, P.~Fritzsch, T.~Korzec, A.~Ramos, S.~Sint, et~al., {\it {Towards a
  new determination of the QCD Lambda parameter from running couplings in the
  three-flavour theory}},  {\em PoS} {\bf LATTICE2014} (2014) 291,
  [\href{http://xxx.lanl.gov/abs/1411.7648}{{\tt arXiv:1411.7648}}].

\bibitem{Fodor:2014cpa}
Z.~Fodor, K.~Holland, J.~Kuti, S.~Mondal, D.~Nogradi, et~al., {\it {The lattice
  gradient flow at tree-level and its improvement}},  {\em JHEP} {\bf 1409}
  (2014) 018, [\href{http://xxx.lanl.gov/abs/1406.0827}{{\tt
  arXiv:1406.0827}}].

\bibitem{Ramos:2014kka}
A.~Ramos and S.~Sint, {\it {On $\mathcal O(a^2)$ effects in gradient flow
  observables}},  \href{http://xxx.lanl.gov/abs/1411.6706}{{\tt
  arXiv:1411.6706}}.

\bibitem{Lin:2014fxa}
C.~J.~D. Lin, K.~Ogawa, H.~Ohki, A.~Ramos, and E.~Shintani, {\it {SU(3) gauge
  theory with 12 flavours in a twisted box}},
  \href{http://xxx.lanl.gov/abs/1410.8824}{{\tt arXiv:1410.8824}}.

\bibitem{Perez:2014isa}
M.~García~Pérez, A.~González-Arroyo, L.~Keegan, and M.~Okawa, {\it {The
  $SU(\infty)$ twisted gradient flow running coupling}},  {\em JHEP} {\bf 1501}
  (2015) 038, [\href{http://xxx.lanl.gov/abs/1412.0941}{{\tt
  arXiv:1412.0941}}].

\bibitem{Cheng:2014jba}
A.~Cheng, A.~Hasenfratz, Y.~Liu, G.~Petropoulos, and D.~Schaich, {\it
  {Improving the continuum limit of gradient flow step scaling}},  {\em JHEP}
  {\bf 1405} (2014) 137, [\href{http://xxx.lanl.gov/abs/1404.0984}{{\tt
  arXiv:1404.0984}}].

\bibitem{Hasenfratz:2014rna}
A.~Hasenfratz, D.~Schaich, and A.~Veernala, {\it {Nonperturbative beta function
  of eight-flavor SU(3) gauge theory}},
  \href{http://xxx.lanl.gov/abs/1410.5886}{{\tt arXiv:1410.5886}}.

\bibitem{Fodor:2015baa}
Z.~Fodor, K.~Holland, J.~Kuti, S.~Mondal, D.~Nogradi, et~al., {\it {The running
  coupling of 8 flavors and 3 colors}},
  \href{http://xxx.lanl.gov/abs/1503.0113}{{\tt arXiv:1503.0113}}.

\bibitem{Aoki:2014latt}
Y.~Aoki, {\it {Near conformal strong dynamics}},  {\em Talk at The 32nd
  International Symposium on Lattice Field Theory} (2014).

\bibitem{Rantaharju:2013bva}
J.~Rantaharju, {\it {The Gradient Flow Coupling in Minimal Walking
  Technicolor}},  {\em PoS} {\bf Lattice2013} (2014) 084,
  [\href{http://xxx.lanl.gov/abs/1311.3719}{{\tt arXiv:1311.3719}}].

\bibitem{Perez:2014sqa}
M.~G. Perez, A.~Gonzalez-Arroyo, and M.~Okawa, {\it {Volume independence for
  Yang-Mills fields on the twisted torus}},
  \href{http://xxx.lanl.gov/abs/1406.5655}{{\tt arXiv:1406.5655}}.

\bibitem{Shindler2014}
A.~Shindler, {\it {Chiral Ward identities, automatic O(a) improvement and the
  gradient flow}},  {\em Nucl.Phys.} {\bf B881} (2014) 71--90,
  [\href{http://xxx.lanl.gov/abs/1312.4908}{{\tt arXiv:1312.4908}}].

\bibitem{DelDebbio:2013zaa}
L.~Del~Debbio, A.~Patella, and A.~Rago, {\it {Space-time symmetries and the
  Yang-Mills gradient flow}},  {\em JHEP} {\bf 1311} (2013) 212,
  [\href{http://xxx.lanl.gov/abs/1306.1173}{{\tt arXiv:1306.1173}}].

\bibitem{Suzuki:2013gza}
H.~Suzuki, {\it {Energy-momentum tensor from the Yang-Mills gradient flow}},
  {\em PTEP} {\bf 2013} (2013), no.~8 083B03,
  [\href{http://xxx.lanl.gov/abs/1304.0533}{{\tt arXiv:1304.0533}}].

\bibitem{patellaconf}
A.~Patella, {\it {Energy-momentum tensor on the lattice and Wilson flow}.},
  {\em Talk at The 32nd International Symposium on Lattice Field Theory}
  (2014).

\bibitem{Shindler:2014oha}
A.~Shindler, J.~de~Vries, and T.~Luu, {\it {Beyond-the-Standard-Model matrix
  elements with the gradient flow}},  {\em PoS} {\bf LATTICE2014} (2014) 251,
  [\href{http://xxx.lanl.gov/abs/1409.2735}{{\tt arXiv:1409.2735}}].

\bibitem{Brida:2013mva}
M.~Dalla~Brida and D.~Hesse, {\it {Numerical Stochastic Perturbation Theory and
  the Gradient Flow}},  {\em PoS} {\bf Lattice2013} (2014) 326,
  [\href{http://xxx.lanl.gov/abs/1311.3936}{{\tt arXiv:1311.3936}}].

\bibitem{Monahan:2015lha}
C.~Monahan and K.~Orginos, {\it {Locally smeared operator product expansions in
  scalar field theory}},  \href{http://xxx.lanl.gov/abs/1501.0534}{{\tt
  arXiv:1501.0534}}.

\end{thebibliography}\endgroup

\end{document}